\documentclass[useAMS,usenatbib,usegraphicx]{mn2e}

\newcommand{\asec}{$^{\prime\prime}$}

\def\H{N$_{2}$H$^{+}$}
\def\D{N$_{2}$D$^{+}$}

\def\AMM{NH$_3$}

\def\CII{\mbox{C$^{18}$O}}

\def\HII{H{\sc ii}}

\def\kms{\mbox{km~s$^{-1}$}}
\def\cmc{cm$^{-3}$}
\def\cmq{cm$^{-2}$}

\def\solm{\mbox{M$_\odot$}}

\def\Vlsr{$V_{\rm LSR}$}
\def\Dfrac{$D_{\rm frac}$}

\def\Tex{\mbox{$T_{\rm ex}$}}

\def\Tk{\mbox{$T_{\rm k}$}}
\def\Tr{\mbox{$T_{\rm rot}$}}

\def\mvir{\mbox{$M_{\rm VIR}$}}

\title[CO depletion in infrared dark clouds]{High CO depletion in southern infrared-dark clouds}
\author[F. Fontani et al.]{F. Fontani$^{1}$\thanks{E-mail:
fontani@arcetri.astro.it }, A. Giannetti$^{2}$, M.T. Beltr\'an$^{1}$, R. Dodson$^{3}$, M. Rioja$^{3}$, J. Brand$^{2}$
\newauthor
P. Caselli$^{4}$ and R. Cesaroni$^{1}$
\\
\\
$^{1}$ INAF-Osservatorio Astrofisico di Arcetri, L.go E. Fermi 5, Firenze, I-50125, Italy \\
$^{2}$ INAF-Istituto di Radioastronomia, via P. Gobetti 101, Bologna, I-40129, Italy \\
$^{3}$ International Centre for Radio Astronomy Research, University of Western Australia, Perth, Australia \\
$^{4}$ School of Physics and Astronomy, University of Leeds, Leeds, LS2 9JT , UK 
}

\begin{document}

\date{Accepted date. Received date; in original form date}

\pagerange{\pageref{firstpage}--\pageref{lastpage}} \pubyear{2011}

\maketitle

\label{firstpage}

\begin{abstract}
Infrared-dark high-mass clumps are among the most promising objects to study
the initial conditions of the formation process of high-mass stars and
rich stellar clusters. In this work,
we have observed the (3--2) rotational transition of \CII\ with the APEX telescope, 
and the (1,1) and (2,2) inversion transitions of \AMM\ with
the Australia Telescope Compact Array in 21 infrared-dark clouds 
already mapped in the 1.2~mm continuum, with the aim of measuring basic chemical
and physical parameters such as the CO depletion factor ($f_{\rm D}$), the gas kinetic
temperature and the gas mass. In particular, the \CII\ (3--2) line allows us to derive
$f_{\rm D}$ in gas at densities higher (and hence potentially more depleted)
than that traced by the (1--0) and (2--1) lines, typically used in previous works.
We have detected \AMM\ and \CII\ in all targets.
The clumps have a median mass of $\sim 244$ \solm , are
gravitationally bound, have an average kinetic temperature of 17~K 
and possess mass, H$_2$ column and surface densities consistent 
with being potentially the birthplace of high-mass stars.
We have measured $f_{\rm D}$ in between 5 and 78, with a mean value
of 32 and a median of 29. These values are, to our knowledge, larger than
the typical CO depletion factors measured towards infrared-dark clouds and
high-mass dense cores, and are comparable to or larger than the values measured in
low-mass pre--stellar cores close to the onset of the gravitational collapse.
This result suggests that the earliest phases of the high-mass star and stellar
cluster formation process are characterised by $f_{\rm D}$ larger than 
in low-mass pre--stellar cores.
On the other hand, $f_{\rm D}$ does not seem to be correlated to any
other physical parameter, except for a faint anti-correlation with the
gas kinetic temperature.
Thirteen out of 21 clumps are undetected in the 24 $\mu$m Spitzer images,
and have slightly lower kinetic temperatures, masses and
H$_2$ column densities with respect to the eight Spitzer-bright sources.
This could indicate that the Spitzer-dark clumps are either less evolved 
or are going to form less massive objects. 
\end{abstract}

\begin{keywords}
Molecular data -- Stars: formation -- radio lines: ISM -- submillimetre: ISM -- ISM: molecules 
\end{keywords}

\section{Introduction}
\label{intro}

An ever increasing number of observational evidences indicates that
the earliest phases of massive star and stellar cluster formation occur
within infrared dark clouds (IRDCs). These are dense molecular clouds 
seen as extinction features against the bright mid-infrared Galactic background
(e.g. Simon et al~\citeyear{simon}, Ragan et al.~\citeyear{ragan06}, ~\citeyear{ragan11},
Rathborne et al.~\citeyear{rathborne07},~\citeyear{rathborne10}, Butler \& Tan~\citeyear{bet}). 
IRDCs are characterised by very high
gas column densities ($10^{23} - 10^{25}$\cmc ) and low temperatures
($\leq 25~K$), so that they are believed to be the place where most
of the stars in our Galaxy are being formed. 

From an observational point of view,
the spatial distribution of the IRDCs in the Galaxy follows that of the 
molecular galactic component (with a concentration in the so-called 5~kpc 
molecular ring), and they are strong emitters of both far-IR/millimetre 
continuum and rotational molecular transitions, especially those characterised
by a high critical density. In many of them,
observations of the dense gas at sub-parsec linear scales
revealed the presence of on-going star formation, both in isolated
and clustered mode (e.g.~Beuther \& Sridharan~\citeyear{bes},~Zhang et al.~\citeyear{zhang09}, 
Fontani et al.~\citeyear{fonta09}, Jim\'enez-Serra et al.~\citeyear{jimenez},
Pillai et al.~\citeyear{pillai11}), including clear signs 
of high-mass star formation like hot cores (e.g.~Rathborne et al.~\citeyear{rathborne08})
and/or Ultracompact \HII\ regions (Battersby et al.~\citeyear{battersby}).
This demonstrates that IRDCs are indeed the birthplace of stars and stellar 
clusters of all masses. Therefore, the IRDCs in the earliest evolutionary stages 
are the best locations where to 
study the initial conditions of the star formation process and put constraints 
on current theories.

Despite the identification of thousands of IRDCs, the number of
studies devoted to unveiling their physical and chemical properties 
remains still limited. There are very few targeted studies, especially in the southern 
hemisphere (e.g. Vasyunina et al.~\citeyear{vasyunina09},
Miettinen et al.~\citeyear{miettinen}), where there are fewer groundbased facilities operating
in the (sub-)millimetre and centimetre domain than there are in the 
north. In particular, very little
is known about the chemistry of IRDCs: studies
suggest chemical compositions similar to those observed in low-mass pre--stellar
cores (Vasyunina et al.~\citeyear{vasyunina09}), including large 
abundances of deuterated species (Pillai et al.~\citeyear{pillai07}; Pillai
et al.~\citeyear{pillai11}; Fontani et al.~\citeyear{fonta11}). However, the amount of CO freeze-out, 
a key chemical parameter for pre--stellar cores 
(see e.g. Bergin \& Tafalla~\citeyear{bet07} for a review) remains
controversial: some works indicate high levels (factor of 5, 
Hern\'andez et al.~\citeyear{hernandez}) of CO freeze-out,
while others do not reveal significant CO
depletion (e.g. Miettinen et al.~\citeyear{miettinen}).

In this work we present observations of the rotational transition
(3--2) of the dense gas tracer \CII , performed with the Atacama Pathfinder 
EXperiment (APEX) 12-m Telescope, 
and of the inversion transitions (1,1) and (2,2) of \AMM\ carried out
with the Australia Telescope Compact Array (ATCA), towards 21 IRDCs with declination
lower than $-30^{\circ }$ already mapped in the 1.2~mm continuum. 
The \CII\ observations allow to compute the
CO depletion factor, while the ammonia inversion transitions can be used 
to derive the temperature in dense and cold gas (see e.g.~Ho \& Townes~\citeyear{het}).
Observations of \H\ and \D\ in some of the target sources, 
useful to derive the amount of deuterated fraction (by comparing
the \D\ and \H\ column densities), are also presented.
In Sect.~\ref{selection} we describe the criteria applied to select the
targets. Sect.~\ref{obs} gives an overview of the observations.
The results are presented in Sect.~\ref{res} and discussed in Sect.~\ref{discu}.
Conclusions and a summary of the main findings are given in Sect.~\ref{conc}.

\section[]{Target selection}
\label{selection}

We selected 21 IRDCs from the 95 massive millimetre clumps 
detected by Beltr\'an et al.~(2006) in the 1.2~mm continuum and 
non-MSX emitters (neither diffuse nor point-like). 
The targets were chosen according to 
these criteria: (i) source declination $\delta \leq -30^{\circ }$;
(ii) clumps isolated or having the emission peak separated by more than 
the SIMBA half power beam width to that of MSX-emitter objects, to limit 
confusion and select the most quiescent sources; 
(iii) clump masses above $\sim 35 M_{\odot}$ to deal with possible massive 
star formation.
In this work we have recomputed the masses
utilising the gas temperature derived from ammonia for each clump,
and our results confirm that all targets are high-mass clumps
(see Sect~\ref{mass}).
The list of IRDCs is given in Table~\ref{sources}.
The coordinates correspond to the peak of the 1.2~millimetre continuum
emission mapped by Beltr\'an et al.~(\citeyear{beltran06}). 
In Table~\ref{sources} we also give some basic information like the distance 
to the Sun, the Galactocentric distance, and the (non-)detection in the Spitzer-MIPS
24 $\mu$m images. A comparison between the 1.2~mm continuum maps and the
Spitzer 24 $\mu$m images of each target (except for 13039--6108c6, for which
the MIPS images are not available) is shown in Fig.~\ref{appA-fig1} of 
Appendix~\ref{appendixA}. 

\section{Observations}
\label{obs}

\subsection{APEX}
\label{apex}

Single-point spectra of the \CII\ (3--2), \H\ (3--2) and \D\ (4--3) lines 
towards the sources listed in Table~\ref{sources} were obtained
with the APEX Telescope in service mode
between the 20th and the 28th of June, 2008. The observations were performed 
in the wobbler-switching mode with a 150\arcsec\ azimuthal throw and a chopping 
rate of 0.5 Hz. The receiver used for all lines was SHFI/APEX-2. 
The backend provided a total bandwidth of 1000 MHz. 
Details about the
observed lines and the observational parameters are given in Table~\ref{obs_apex}.
The telescope pointing and focusing were checked regularly by continuum scans on 
planets and the corrections were applied on-line. 
Calibration was done by the chopper-wheel technique.
Spectra were obtained in antenna temperature units (corrected for atmospheric
attenuation), $T_A^*$, and then converted to main beam temperature units through the
relation $T_{\rm MB}=\frac{F_{\rm eff}}{B_{\rm eff}} T_A^*$ ($F_{\rm eff}$ and $B_{\rm eff}$
are given in Col.~5 of Table~\ref{obs_apex}). 
The velocities adopted to centre the backends are given in Col.~6 of Table~\ref{sources}.
For most sources we knew the radial Local Standard of Rest velocities ($V_{\rm LSR}$)
from previous CS observations. The spectra of 16435--4515c3 (for which
we did not have the CS data) were centred 
at $V_{\rm LSR} = 0$; in any case the total bandwidth of the backend 
($\sim 1000$ \kms ) was larger than the velocity gradient across the Galaxy. 

\begin{table*}
 \centering
 \begin{minipage}{180mm}
  \caption{Source list and detection summary.}
  \label{sources}
  \begin{tabular}{cccccclcl}
  \hline
  Source name$^{a}$     &     R.A. (J2000)       &  Dec. (J2000) & $l$ & $b$ & $V_{\rm LSR}$$^{b}$ & $d$$^{c}$ & $D_{\rm GC}$ & MIPS 24 $\mu$m$^{d}$ \\
    & h m s  & $^o\,\, \arcmin \,\, \arcsec$  & $^o$  & $^o$  & \kms\ & kpc & kpc &   \\
  \hline
08477--4359c1 &  08:49:35.13 & --44:11:59 &   264.69 &  --0.07 &   8.5 & 1.8 & 8.9 & Y  \\
13039--6108c6  & 13:07:14.80 & --61:22:55   & 305.18 &   1.14 & --26.2 & 2.4 & 7.4 & --  \\
13560--6133c2  & 13:59:33.04 & --61:49:13 &   311.23 &  --0.35  &--58.4 & 5.6 & 6.4 &N \\
14183--6050c3  & 14:22:21.54 & --61:06:42  & 314.03  & --0.52 & --42.6 & 3.4 & 6.6 &N \\
15038--5828c1  & 15:07:32.52 & --58:40:33  &  320.19  & --0.77 & --67.1 & 5.0 & 5.7 & N \\
15278--5620c2  & 15:31:44.17 & --56:32:08  &  324.07  & --0.73 & --49.4 & 3.4 & 6.1 &N \\
15470--5419c1   & 15:51:28.24 & --54:31:42 &   327.51  & --0.83 & --61.7 & 4.1 & 5.5 &Y$^{e}$ \\
15470--5419c3   & 15:51:01.62 & --54:26:46 &   327.51  & --0.72 & --61.7 & 4.1 & 5.5 & Y \\
15557--5215c2   & 15:59:36.20 & --52:22:58  &  329.81  &  0.03 & --67.6 & 4.4 & 5.2 & Y \\
15557--5215c3   & 15:59:39.70 & --52:25:14  &  329.80  & --0.00 & --67.6 & 4.4 & 5.2 & N$^{f}$  \\
16061--5048c1   & 16:10:06.61 & --50:50:29  &  332.06  &  0.08 & --51.8 & 3.6 & 5.6 & Y  \\
16061--5048c4   & 16:10:06.61 & --50:57:09  &  331.98  &  0.00 & --51.8 & 3.6 & 5.6 & N$^{f}$ \\
16093--5128c2   & 16:12:55.46 & --51:43:22  &  331.77  & --0.86 & --97.3 & 6.1 & 4.3 & N  \\
16093--5128c8   & 16:12:49.63 & --51:36:34  &  331.84  & --0.77 & --97.3 & 6.1 & 4.3 & N$^{g}$  \\
16164--4929c3   & 16:20:24.51& --49:35:34  &  334.11  & --0.17 & --33.8 & 2.6 & 6.3 & N$^{g}$  \\
16254--4844c1   & 16:29:00.89 & --48:50:31  &  335.63 &  --0.65 & --45.5  & 3.4 & 5.6 & Y \\
16435--4515c3   & 16:47:33.13 & --45:22:51  &  340.31  & --0.71  &  0.0   & --$^{h}$ & 4.9 & N  \\
16482--4443c2   & 16:51:44.59 & --44:46:50  &  341.24  & --0.90 & --43.3  & 3.7 & 5.1 & N$^{f}$  \\ 
16573--4214c2   & 17:00:33.38 & --42:25:18  &  344.08  & --0.67 & --23.7  & 2.6 & 6.0 & Y  \\
17040--3959c1   & 17:07:58.78 & --40:02:24   & 346.82  & --0.35 &  --0.4  & 16.5$^{i}$ & 8.5 & N?$^{l}$ \\
17195--3811c2   & 17:23:00.30 & --38:14:58  &  349.97  & --1.68 & --26.4  & 3.6 & 5.0 & Y \\
\hline
\end{tabular}

$^{a}$ derived from the IRAS source name and the 'clump' number assigned by Beltr\'an et al~(\citeyear{beltran06}); \\
$^{b}$ from the CS (2--1) or (3--2) transitions (Fontani et al.~\citeyear{fonta05}); \\
$^{c}$ kinematic distance from the Sun (Beltr\'an et al~\citeyear{beltran06}). In case of distance ambiguity, the near distance is adopted; \\
$^{d}$ detection (Y) or non-detection (N) of the clump in the Spitzer-MIPS 24$\mu$m band; \\
$^{e}$ point-like infrared source not at clump centre, probably associated with it; \\
$^{f}$ point-like infrared source at the edge of the 3 $\sigma$ rms contour level of the millimetre continuum, probably not associated with it; \\
$^{g}$ diffuse infrared emission inside the 3 $\sigma$ rms contour level of the millimetre continuum, probably associated with a nearby infrared object; \\
$^{h}$ the kinematic distance recomputed in Sect.~\ref{mass} from the \CII\ (3--2) line peak velocity is 3.1~kpc; \\
$^{i}$ near distance smaller than 100~pc, so that the far distance is adopted; \\
$^{l}$ the millimetre map of the clump is incomplete because at the edge of the SIMBA field of view, thus one cannot conclusively assert if the clump is associated with an infrared source (see Fig.~\ref{appA-fig2}).  \\
\end{minipage}
\end{table*}

\begin{table*}
\centering
\begin{minipage}{140mm}
\caption{Transitions observed with APEX and observational parameters.}
\label{obs_apex}
\begin{tabular}{cccccc}
\hline
Transition & $\nu$$^{a}$ & HPBW & $\Delta V$$^{b}$ & $B_{\rm eff}/F_{\rm eff}$ & $T_{\rm sys}$ \\
  & GHz & \arcsec\ & \kms\ & & K \\
\hline
\CII\ (3--2) & 329.331 & 18.9 & $\sim 0.22$ & 0.74/0.95 & 200 -- 300$^{c}$ \\
  \D\ (4--3) & 308.422$^{d}$ & 20.2 & $\sim 0.23$ & 0.74/0.95 & 160 -- 300  \\
  \H\ (3--2) & 279.512$^{e}$ & 22.3 & $\sim 0.26$ & 0.74/0.95 & 130 -- 160 \\
\hline
\end{tabular}

$^{a}$ line rest frequency; \\
$^{b}$ spectral resolution; \\
$^{c}$ for 13039--6108c6, 13560--6133c2 and 14183--6050c3, $T_{\rm sys}\sim 600$ K; \\
$^{d}$ from the Cologne Database for Molecular Spectroscopy (CDMS), and it refers to the $F_{1} F = 5\;5\rightarrow 4\;4$ 
hyperfine component; \\
$^{e}$ from Pagani et al.~(\citeyear{pagani}), and it refers to the $F_{1} F = 4\;5\rightarrow 3\;4$ 
hyperfine component, which has a relative intensity of $17.46\%$. \\
\end{minipage}
\end{table*}

\subsection{ATCA}
\label{atca}

We observed the inversion transitions $(J,K)=$ (1,1) and (2,2) of ammonia
at  23694.5~MHz and 23722.6~MHz  (K-band at $\sim 1.2$~cm), respectively,
with the ATCA\footnote{The Australia Telescope Compact Array is part of the Australia Telescope 
which is funded by the Commonwealth of Australia for operation as a National 
Facility managed by CSIRO.} towards all targets in Table~\ref{sources}. 
The observations were performed 
between the 4th and the 8th of March, 2011, for a total telescope time of 48 hours. 
We used the configuration 750D, which provides baselines between 31 and
4469 m. The primary beam was $\sim 2.5$\arcmin\ at the line frequencies.
The flux density scale was established by observing the standard primary calibrator
1934--638, and the uncertainty is expected to be of the order of $\sim 10\%$. 
Gain calibration was ensured by frequent observations of nearby compact quasars.
The quasar 0537--441 was used for passband calibration.
Pointing corrections were derived from nearby quasars and applied online.
Atmospheric conditions were generally good (weather path noise $\simeq 400\;\mu$m
or better).

The total time was broken up into series of 3-/5-minute snapshots in order to
improve the coverage of the uv-plane for each target.
As a consequence of this observing strategy, the integration time on source is
variable, and generally in between $\sim 30$ mins and $\sim 1$ hour.
The CABB\footnote{http://www.narrabri.atnf.csiro.au/observing/CABB.html} 
correlator provided two "zoom" bands of 64~MHz each, with a 
spectral resolution in each zoom of 32 kHz ($\sim $0.4 \kms\
at the frequencies of the lines). 
The ammonia lines were observed in one zoom band.
The data were edited and calibrated following standard tasks and procedures
of the MIRIAD software package. 
After the editing and calibration in MIRIAD, the data were imported in AIPS.
Imaging and deconvolution were performed using the 'imagr' task,
applying natural weighting to the visibilities. The ammonia emission was
detected only on the short baselines, thus we discarded all baselines $>
30\,\mathrm{k}\lambda$. In order to obtain images with the same angular
resolution, and comparable to the APEX beam, we reconstructed the images 
with a circular beam of $20\arcsec$ of diameter for all sources, 
except for 17195-3811c2 and 17040-3959c1, 
that have a poorer UV-coverage, and for which we get a beam of roughly 
$20\arcsec \times 40\arcsec$. 
Moreover, 16254-4844c1 and 16573-4214c2 were observed only once, making 
the clean impossible and thus forcing us to use the dirty image deconvolved
with the dirty beam to determine the spectrum.

In this work the ATCA data will be used only to derive the gas temperature
from the \AMM\ (1,1) and (2,2) spectra at the dust emission peak (in Sect.~\ref{trot}). 
A complete presentation and analysis of the ATCA observations
will be given in a forthcoming paper (Giannetti et al., in prep).




\section{Results}
\label{res}

\subsection{Detection summary}
\label{summary}

The detection summary is given in Table~\ref{detection}.
\CII\ (3--2) was observed and detected in all sources. Twelve targets were also
observed and detected in \H\ (3--2). Ten clumps were observed
in \D\ (4--3), and only clump 15557--5215c2 was marginally detected.
This is, to our knowledge, the first detection of this line towards an
IRDC.
All targets were observed and detected in the ammonia inversion transitions, with
the exception of 14183--6050c3 which was undetected in the (2,2) line.

\begin{table*}
 \centering
 \begin{minipage}{160mm}
  \caption{Detection summary: Y = detected, N = not detected, -- = not observed;}
  \label{detection}
  \begin{tabular}{ccccc}
  \hline
Source   &  & APEX &  & ATCA (\AMM )  \\
 \cline{2-4}
       &  \CII\ (3--2) & \H\ (3--2) & \D\ (4--3) & \AMM\ (1,1) / (2,2) \\
  \hline
08477--4359c1  & Y & Y & N & Y / Y \\
13039--6108c6  & Y & Y & N &   Y / Y  \\
13560--6133c2  & Y & --  & -- &  Y / Y  \\
14183--6050c3  & Y & -- & -- & Y / N  \\
15038--5828c1  & Y & -- & N &   Y / Y  \\
15278--5620c2  & Y & --& N &  Y / Y  \\
15470--5419c1  & Y & Y & N & Y / Y  \\
15470--5419c3  & Y & Y & N &  Y / Y  \\
15557--5215c2  & Y & Y & Y$^{a}$ & Y / Y  \\
15557--5215c3  & Y & Y & N & Y / Y  \\
16061--5048c1  & Y & Y & N &  Y / Y  \\
16061--5048c4  & Y & Y & N & Y / Y  \\
16093--5128c2  & Y & Y & -- & Y / Y  \\
16093--5128c8  & Y & -- & -- & Y / Y  \\
16164--4929c3  & Y & -- & -- & Y / Y  \\
16254--4844c1  & Y & -- & -- & Y / Y  \\
16435--4515c3  & Y & Y & -- & Y / Y  \\
16482--4443c2  & Y & Y & -- & Y / Y  \\
16573--4214c2  & Y & Y & -- & Y / Y   \\
17040--3959c1  & Y & -- & -- & Y / Y  \\
17195--3811c2  & Y & -- & -- & Y / Y  \\
\hline
\end{tabular}

$^{a}$ marginal detection (see Sect.~\ref{n2dpcoldens}). \\
\end{minipage}
\end{table*}

\subsection{Line profiles and parameters}
\label{profiles}

All spectra of the \CII\ and \H\ lines are shown in Appendix~\ref{spectra}. 
Fig.~\ref{spectra_co_n2hp}
shows the \H\ (3--2) and \CII\ (3--2) transitions in the twelve clumps
observed and detected in both lines. Fig.~\ref{spectra_co_only} shows the
nine sources observed and detected in \CII\ (3--2) only.

\begin{figure}
\begin{minipage}{80mm}
 \begin{center}
 \resizebox{\hsize}{!}{\includegraphics[angle=-90]{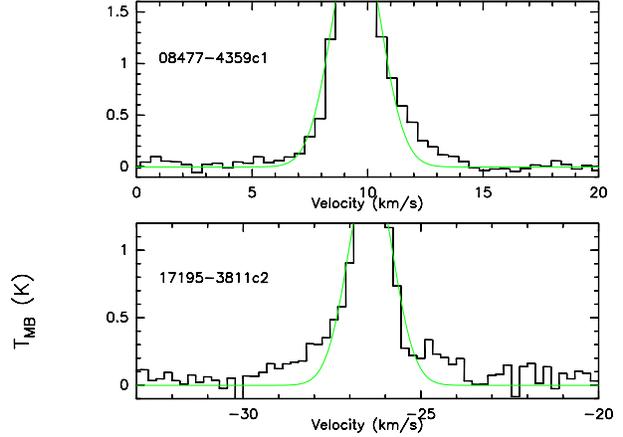}}
 \caption[]
 {\label{spec_fits}{\CII\ (3--2) spectra of 08477--4359c1 and
17195--3811c2. The Gaussian fits are superimposed on the
spectra and highlight significant emission in non-Gaussian wings.}}
 \end{center}
 \end{minipage}
\end{figure}

Most of the \CII\ lines are well fitted by single Gaussians.
The results of these fits are reported in Table~\ref{tab_lin_c18o}.
Slightly asymmetric profiles deviating from the Gaussian shape can be 
noted in 15470--5419c1, 16061--5048c1, 
16482--4443c2, 16573--4214c2 and 17040--3959c1. This could be due either to the superposition 
of multiple blended velocity components, or to high optical depth effects.
Non-Gaussian wings in one (or both) side(s) of the line indicating presence 
of outflows are noticeable towards two sources, 08477--4359c1 and
17195--3811c2 (see Fig.~\ref{spec_fits}).
Four sources show multiple lines well-separated in velocity, 
14183--6050c3, 16093--5128c2, 16093--5128c8, 16435--4515c3.
Because we had the previous CS observations (Fontani et al.~\citeyear{fonta05})
as reference, in Table~\ref{tab_lin_c18o} we have identified the line associated with the
star-forming region of our interest. The other components are almost
certainly arising from clouds along the line of sight but not associated
with the star-forming region, given the large separation in velocity (from $\sim 20$ to $\sim 40$ \kms ,
see Figs.~\ref{spectra_co_n2hp} and \ref{spectra_co_only}).

\begin{table}
\centering
\begin{minipage}{90mm}
\caption{\CII\ line parameters derived from Gaussian fits. For sources with more than
one velocity component, the line labelled as '--a' is the one associated with the IRDC (based
on previous CS observations, Fontani et al.~\citeyear{fonta05}).}
 \label{tab_lin_c18o}
\begin{tabular}{lcccc}
\hline
source & $\int T_{\rm MB}{\rm d}v$ & $v_{\rm LSR}$ & FWHM & $T_{\rm peak}$ \\
   &   K \kms\  & \kms\ & \kms\ & K \\
   \hline
        08477--4359c1  &     6.10(0.09) & 9.53(0.02) &      2.44(0.04) &      2.3 \\ 
       13039--6108c6 &       1.4(0.1)  & -26.04(0.06) &      1.8(0.1)  &      0.76  \\
       13560--6133c2  &      1.6(0.15) & -57.47(0.08) &      1.9(0.2) &   0.8  \\
       14183--6050c3--a &     0.9(0.1)  & -45.52(0.08) &      1.5(0.2) &	      0.56  \\
       14183--6050c3--b &     0.57(0.06) & -26.70(0.02) &     0.48(0.05) &      1.1 \\  
    15038--5828c1 &         0.9(0.1) & 	-69.3(0.3) &	    5.2(0.7) &	      0.16 \\ 
  15278--5620c2 &            2.5(0.1) & 	-49.42(0.04) &      2.0(0.1) &	       1.18  \\
  15470--5419c1     &        2.3(0.1) & 	-58.74(0.07) &      2.45(0.2) &	      0.88  \\
  15470--5419c3     &        2.6(0.1) & 	-61.38(0.06) &      2.9(0.1) &	      0.84  \\
  15557--5215c2     &        4.2(0.1) & 	-67.57(0.03) &      2.28(0.08) &      1.7  \\
  15557--5215c3     &        1.7(0.1) & 	-68.47(0.08) &      2.4(0.2) &	      0.67  \\
  16061--5048c1     &        7.61(0.02) &	-67.28(0.03) &     3.58(0.06) &      2.0  \\
  16061--5048c4     &        3.2(0.1)  & 	-51.99(0.03) &      1.94(0.07) &      1.6  \\
  16093--5128c2--a     &       3.6(0.1)  & 	-96.19(0.05) &      3.4(0.1)  &	       1.0 \\ 
  16093--5128c2--b     &       1.97(0.07) &	-63.46(0.03) &      1.57(0.07) &      1.2  \\
  16093--5128c8--a     &      0.96(0.06) &	-96.85(0.08) &      2.3(0.2) &	      0.39  \\
  16093--5128c8--b 	&     0.19(0.04)  &	-68.9(0.1) &	   1.0(0.2) &	      0.18  \\
       16164--4929c3  &      2.95(0.07) &	-33.80(0.02) &      1.76(0.05) &      1.6  \\
       16254--4844c1  &      2.72(0.09) &	-44.98(0.03) &       2.20(0.09) &      1.2  \\
       16435--4515c3--a   &     1.7(0.1) &	-35.57(0.04) &      1.6(0.1) &	       1.0 \\ 
      16435--4515c3--b &    0.88(0.09) &	-52.09(0.06) &      1.3(0.2)	&      0.62  \\
      16435--4515c3--c   &    0.5(0.1)  & 	-24.4(0.1)  &	    1.5(0.4) &	      0.31  \\
       16482--4443c2  &      4.6(0.1) &    -43.35(0.05)  &      3.3(0.1)	&       1.3 \\ 
       16573--4214c2  &      8.1(0.2) & 	-27.29(0.03) &      2.76(0.06) &      2.8 \\ 
       17040--3959c1  &      6.45(0.08) &	-15.87(0.02) &    2.40(0.03) &     2.5  \\
       17195--3811c2  &      2.54(0.07) &	-26.41(0.02) &      1.47(0.06) &      1.6  \\
   \hline
 \end{tabular}
 \end{minipage}
 \end{table}

The \H\ (3--2) lines roughly peak at the same position as the \CII\ (3--2) ones and have comparable
line widths. Because of the hyperfine structure present in the transition, it is not straightforward to
derive conclusions on blended multiple velocity components. However, we note hints of possible 
secondary velocity components in all the lines that show similar features in the \CII\ (3--2) line 
(15470--5419c1, 16061--5048c1, 16573--4214c2) except for 16482--4443c2.

\begin{table*}
\centering
\begin{minipage}{160mm}
\caption{\H\ line parameters and column densities. For optically thin transitions 
without well-constrained opacity ($\tau=0.1$), to compute $N$(\H ) 
we adopted a \Tex = 6~K (see text). For sources with more than
one velocity component, the line labelled as '--a' is the one having the 
peak velocity consistent with that of the clump of our interest,
as in Table~\ref{tab_lin_c18o}.}
\label{tab_lin_n2h}
\begin{tabular}{cccccccc}
\hline
source & vel. range  & $\int T_{\rm MB}{\rm d}v$ & $v_{\rm LSR}$ & FWHM & $\tau$ & \Tex\ & $N$(\H ) \\
   &  \kms\  &   K \kms\  & \kms\ & \kms\ & & K & ($\times 10^{13}$)\cmq\ \\
   \hline
08477--4359c1 &   5.4,14.8 &  6.02(0.07)    &   10.53(0.02)  &    2.87(0.05) & 0.1 & 6 & 2.93(0.03)\\
13039--6108c6 &  --30.0,--23.6 & 0.66(0.08)     & --26.2(0.1)  &    1.6(0.4)	& 0.1 & 6 & 0.32(0.04) \\
15470--5419c1 & --66.1,--54.8 & 5.2(0.1)    & --58.82(0.03) &     2.4(0.2)	& 12(2) & 4.8(0.3) & 54(1) \\
15470--5419c3 & --67.1,--56.8 & 9.4(0.1)     & --61.38(0.03)  &    3.14(0.02)	& 0.1 & 6 & 4.6(0.05) \\
15557--5215c2 & --74.1,--62.6 & 17.2(0.2)     & --66.77(0.02)  &    2.01(0.05)	& 12.7(0.8) & 8.0(0.5) & 27(2) \\
15557--5215c3 & --71.8,--66.2 & 1.12(0.07)     & --68.69(0.06)  &    1.61(0.09)	 & 1.4(0.2) & 4.6(0.2) & 1.8(0.1) \\
16061--5048c1 & --72.6,--59.5 & 25.3(0.2)      & --66.20(0.02)  &    3.14(0.09)	& 9.1(0.8) & 9(1) & 29(3) \\
16061--5048c4 & --56.3,--48.3 & 2.13(0.08)     & --51.98(0.05)  &    1.29(0.09)	& 7(1) & 4.5(0.3) & 18(4) \\
16093--5128c2--a & --102.1,--92.1 & 1.28(0.06)    & --97.20(0.08)  &    3.1(0.2)	& 0.1 & 6 & 0.62(0.03) \\
16093--5128c2--b & --68.0,--60.2 & 0.63(0.06)     & --63.21(0.08)  &    1.7(0.3)	& 0.3(1.5) & 6 & 0.38(0.04) \\
16435--4515c3--a & --53.0,--49.8 & 0.08(0.02) & --51.1(0.3)   &   1.6(0.4)	& 0.1 & 6 & 0.04(0.01)\\
16435--4515c3--b & --37.1,--32.5 & 0.23(0.02)   & --35.2(0.1)   &   1.5(0.3)	& 1(0.2) & 3.5(0.7) & 2.7(0.2) \\ 
16482--4443c2  & --47.0,--38.6  & 3.68(0.07)      & --42.57(0.03)   &   3.02(0.09)	& 0.1 & 6 &  1.79(0.03) \\
16573--4214c2 & --32.5,--22.2 & 9.52(0.08)      & --27.06(0.01)   &   3.79(0.02)	& 0.1 & 6 & 4.64(0.04) \\
 \hline
 \end{tabular}
 \end{minipage}
 \end{table*}

In Table~\ref{tab_lin_n2h} we give the \H\ (3--2) line parameters: 
in \mbox{Cols.~3 -- 7} we list integrated 
intensity ($\int T_{\rm MB}{\rm d}v$), peak velocity ($V_{\rm LSR}$), FWHM,
opacity ($\tau$), and excitation temperature
(\Tex ) of the \H\ (3--2) line, respectively.  Because the rotational transitions of
\H\ possess hyperfine structure, we fitted the lines using the 
METHOD HFS
of the CLASS package\footnote{The CLASS program is
part of the GILDAS software, developed at the IRAM and the Observatoire 
de Grenoble, and is available at http://www.iram.fr/IRAMFR/GILDAS}. 
This method fits all the hyperfine components simultaneously assuming
that they have the same excitation temperature
and width, that the opacity has a Gaussian dependence on frequency and
fixing the separation of the components to the laboratory value.
The line parameters listed in Table~\ref{tab_lin_n2h} have been derived
from this method, except the integrated intensity that has been computed 
by simple integration over the velocity range given in Col.~2. 
The line opacity is deduced from the intensity ratio of the different 
hyperfine components, most of which are blended 
due to the fact that the line widths are larger ($\sim 1 - 3$ \kms )  
than the separation in velocity of the components. However,
the fit residuals are generally low indicating that the procedure provides 
good results despite the blending.


\subsection{Column densities of \H }
\label{n2hpcoldens}

The \H\ column densities were derived following the method outlined in the Appendix A
of Caselli et al.~(\citeyear{caselli02a}). Specifically, we adopted equations (A-1) and (A-4)
for optically thick and optically thin lines, respectively. The method assumes
a constant excitation temperature, \Tex. An estimate of \Tex\ 
can be derived from the fitting procedure to the hyperfine structure
\footnote{See the CLASS user manual for the derivation of
\Tex\ from the output parameters: http://iram.fr/IRAMFR/GILDAS/doc/html/class-html/class.html/},
but not for optically thin lines or lines with opacity not well-constrained.
For these, we have assumed the average \Tex\ ($\sim 6$~K; Table~\ref{tab_lin_n2h})
derived for the sources with well-constrained opacity.

For all sources but 16573--4214c2 and 16482--4443c2, the clump diameter derived from
the 1.2~mm continuum is comparable to or larger than the beam size
(see Table~2 in Beltr\'an et al.~\citeyear{beltran06}, see also Table~\ref{tab_mass}), 
so that we assumed unity filling factor.
This factor was applied also to the unresolved source 16482--4443c2.
We stress that this is an approximation, because the sources could be clumpy. 
However, due to the lack of observations at higher-angular resolution,
neither in the lines observed in this work nor in other high-density
gas tracers, the size of the effective emitting region cannot be determined.
For 16573--4214c2, for which $\theta_{\rm s}\sim 7$\arcsec , 
we applied a correction factor $1/ \eta_{\nu}= (\theta_{\rm s}^2+HPBW^2)/ \theta_{\rm s}^2 \sim 10$. 

The \H\ total column densities are given in Col.~8 of Table~\ref{tab_lin_n2h},
and span a range from $10^{12}$ to  $10^{14}$ \cmq . These values
are in good agreement with those derived from the same line in other
IRDCs (Ragan et al.~\citeyear{ragan06}, Miettinen et al.~\citeyear{miettinen}), 
as well as in other massive starless and star-forming cores 
(Fontani et al.~\citeyear{fonta11}, Chen et al.~\citeyear{chen}).

\subsection{Upper limits on the column density of \D\ and on \D /\H\ column density ratio}
\label{n2dpcoldens}

\begin{figure}
\begin{minipage}{80mm}
 \begin{center}
 \resizebox{\hsize}{!}{\includegraphics[angle=-90]{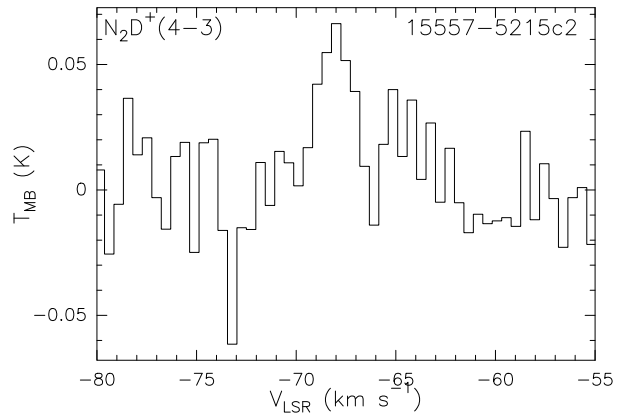}}
 \caption[]
 {\label{spectrum_n2dp43}{\D\ (4--3) spectrum of the only source
 detected in this transition, 15557--5215c2.}}
 \end{center}
 \end{minipage}
\end{figure}

The \D\ (4--3) transition was marginally detected at $\sim 3.2\,\sigma$ rms only 
towards 15557--5215c2. This represents, to our knowledge, the first detection of this line in an IRDC.
The spectrum is shown in Fig.~\ref{spectrum_n2dp43}.
For this source we have computed the \D\ column density fitting
the line with a Gaussian (the results are given in Table~\ref{tab_lin_n2d}), 
and then following the approach of Appendix A in Caselli et al.~(\citeyear{caselli02a}) 
to derive the \D\ column density for the optically thin case. 
This line is also a blend of hyperfine components, therefore the line width
derived from the Gaussian fit is an upper limit to the intrinsic value.
However, because of the poor signal-to-noise ratio of the spectrum,
METHOD HFS did not provide reliable results. As \Tex , we used
that derived from \H\ (3--2) for this source (see Table~\ref{tab_lin_n2h}). 
We find $N$(\D )$\sim 8\times 10^{11}$\cmq\
and the corresponding deuterated fraction (column 
density ratio $N$(\D )/$N$(\H ) = $D_{\rm frac}$) is $\sim 0.003$
(Table~\ref{tab_lin_n2d}). This value is consistent with studies performed in both 
IRDCs (e.g.~Miettinen et al.~\citeyear{miettinen}, Chen et al.~\citeyear{chen})
and massive star-forming clumps containing more evolved objects 
(e.g.~Fontani et al.~\citeyear{fonta06}, \citeyear{fonta11}).

\begin{table*}
\centering
\begin{minipage}{120mm}
\caption{\D\ (4--3) line parameters and \D\ column density in 15557--5215c2}
\label{tab_lin_n2d}
\begin{tabular}{cccccc}
\hline
 $\int T_{\rm MB}{\rm d}v$ & $v_{\rm LSR}$ & FWHM & $T_{\rm peak}$ & $N$(\D ) & $N$(\D )/$N$(\H ) \\
   K \kms\  & \kms\ & \kms\ & K & ($\times 10^{11}$)\cmq\ & \\
   \hline
0.14( 0.03) & --68.1(0.2) &    2.0(0.5) &  0.07 & 8(2)  & 0.003(0.001) \\
 \hline
 \end{tabular}
 \end{minipage}
 \end{table*}

For all the other undetected sources, we have derived
upper limits on the deuterated fraction in between 0.003 and 0.5. 
These upper limits were computed from the integrated intensity upper
limits assuming the lines to be Gaussian 
from the formula $\int T_{\rm MB}{\rm d}v= \frac{\Delta V}{2\sqrt{{\rm ln}2/\pi}}T_{\rm MB}^{\rm peak}$.
We adopted the 3$\sigma$ rms level in the spectrum as peak temperature $T_{\rm MB}^{\rm peak}$,
and the line width measured in the detected source 15557--5215c2 
(Table~\ref{spectrum_n2dp43}) as $\Delta V$. The \Dfrac\ upper limits
are comparable to the values commonly measured in massive clumps 
(Fontani et al.~2006, Chen et al.~\citeyear{chen},
Miettinen et al.~\citeyear{miettinen}), so that our sensitivity does not allow
to conclude if our targets have deuterated fraction similar to or lower than the
other clumps in the literature. We believe that the very low 
detection rate in the \D\ (4--3) line is likely due to the high densities 
required to excite this transition: having a critical density
of $\sim 3 \times 10^{7}$ \cmc , this line is expected to come from very compact 
regions, thus suffering enormous beam dilution effects.

\subsection{Rotation temperatures and \AMM\ column densities at dust emission peak position}
\label{trot}

We have extracted spectra of the \AMM\ (1,1) and (2,2) transitions from the
ATCA channel maps towards the positions given in Table~\ref{sources}. 
The spectra obtained this way were then imported in CLASS, and fitted
using METHOD NH$_3$ to fit the (1,1) lines, which takes into account the
line hyperfine structure, similarly to METHOD HFS (see Sect.~\ref{profiles}). 
This was also used for the (2,2) emission, even though the 
hyperfine components were not always visible, to obtain an upper limit 
for $\tau$.

In Table~\ref{temperature} we list the rotation temperatures (\Tr ), kinetic temperatures (\Tk ),
and total column densities of ammonia ($N_{\rm NH_3}$) derived from the \AMM\ (2,2)/(1,1) line 
ratios, as well as the line peak velocities. We have derived \Tr\ and $N_{\rm NH_3}$ from
the output parameters given by the fitting method outlined above, and using
the formulae given in the Appendix of Busquet et al.~(\citeyear{busquet09}).
The formulae, which result from the discussion in Ho \& Townes (1983, Eq. (4)), have been derived
assuming that the transitions between the metastable inversion doublets are approximated 
as a two-level system, and that the excitation temperature and line width are the same for both 
\AMM\ (1, 1) and \AMM\ (2, 2). Note that the assumption of a two-level system is reasonable 
because transitions between the metastable inversion doublets are usually much faster than 
those of other rotational states (Ho \& Townes~\citeyear{het}). 
\Tk\ was extrapolated from \Tr\ following the empirical approximation 
method described in Tafalla et al.~(\citeyear{tafalla}, see also Eq. (A.5) in Busquet
et al.~\citeyear{busquet09}).
We find kinetic temperatures in between 13 and 25~K, with both mean and median 
\Tk\ of $\sim 17$~K. These values are consistent, within the errors, with typical 
temperatures measured towards IRDCs (Pillai et al.~\citeyear{pillai06}) from
ammonia.

\begin{table}
\centering
 \begin{minipage}{80mm}
  \caption{Rotation temperature, kinetic temperature, total \AMM\ column
  densities and peak velocities derived from the \AMM\ inversion transitions observed 
  with the ATCA. The errors on \Tr\ and \Tk\ are given in parentheses and are computed
  from the propagation of errors. The uncertainty on the column densities includes the error
  on the flux calibration, and is estimated to be of the order of 20 -- 30$\%$. }
  \label{temperature}
\begin{tabular}{lcccc}
  \hline
Source   &  \Tr\ & \Tk\ & $N_{\rm NH_3}$ & $V_{\rm peak}$$^{a}$ \\
 \cline{2-5}
 & & & & \\

  & K & K & $10^{15}$ \cmq\ & \kms\ \\
  \hline
 08477--4359c1 & 16(2)  & 19(3) &	1.06 & 	+11.18(0.02) \\
13560--6133c2 & 15(9) &	17(10)	& 4.61 &	--56.32(0.05) \\
14183--6050c3$^{b}$	& -- &	-- &	-- &	--44.36(0.04) \\ 
15038--5828c1 & 14(3) &	15(4) & 	2.97	& --68.72(0.04) \\
15278--5620c2	& 17(2) &	20(3) & 	3.52 &	--48.69(0.02) \\
15470--5419c1--a &	16(2) & 	18(3) & 	4.23	& --60.72(0.03) \\
15470--5419c1--b &	13(9) & 	14(9) 	& 3.20 &	--57.88(0.06) \\
15470--5419c3 & 17(1) &	19(2) & 3.61	& --61.02(0.01) \\
15557--5215c2	& 19(2) &	23(4) 	& 4.62	& --67.11(0.01) \\
15557--5215c3	& 14(3) &	15(3) 	& 4.47	& --68.53(0.01) \\
16061--5048c1	& 20(3) &	25(5) 	& 4.44	& --66.76(0.01) \\
16061--5048c4	& 12(2) &	13(2)  & 	5.28	& --51.62(0.01) \\
16093--5128c2	& 14(5) &	16(5) &	3.87 & --97.09(0.05) \\
16093--5128c8	& 14(6) &	15(7) & 	1.09	& --95.80(0.04) \\
16164--4929c3	& 13(6) &	14(6) 	& 2.59 &	--32.80(0.01) \\
16254--4844c1$^{c}$ & 18(4) &	22(6)	& 4.77 &	-38.45(0.05) \\ 
16435--4515c3	& 11(4) & 	12(5) & 	1.84 & --51.69(0.04) \\
16482--4443c2	& 15(3) &	16(4) 	& 0.78	& --42.35(0.03) \\	
16573--4214c2$^{c}$ & 15(4) &	17(5) &	3.32	 & -25.09(0.04) \\ 
17040--3959c1	& 16(3) &	18(4) 	& 6.81	& --15.62(0.02) \\
17195--3811c2	& 15(1) &	17(2) & 1.13	& --25.51(0.01) \\
\hline
\end{tabular}

$^{a}$ derived from the (1,1) transition. The uncertainty (between parentheses) is that given by the fitting procedure; \\
$^{b}$ undetected in the (2,2) transition (see Table~\ref{detection}); \\
$^{c}$ observed only for one cycle with bad UV coverage. The values listed have been
derived from the spectra extracted from the dirty map deconvolved with the dirty beam. \\
\end{minipage}
\end{table}

\subsection{H$_2$ masses and other physical parameters from 1.2~mm continuum emission}
\label{mass}

Gas masses were calculated by Beltr\'an et al.~(\citeyear{beltran06}) for all
clumps from the 1.2~mm continuum integrated flux density, assuming optically
thin emission and a reasonable dust temperature of 30~K for all clumps. This 
latter assumption was due to the fact that a temperature estimate for each clump
was lacking. In this work we can profit from the temperatures derived from ammonia 
(Col.~3 of Table~\ref{temperature}) and recompute the H$_2$ masses assuming 
that the dust temperature equals the kinetic temperature. This method implies coupling between 
gas and dust, which is a realistic assumption for gas densities as those of our clumps
($\geq 10^4$ \cmc ).

The gas mass, $M$, has been derived using Eq. (1) in Beltr\'an et al.~(\citeyear{beltran06}), and
adopting the same assumptions, except the dust temperature for which we have
utilised \Tk\ derived from ammonia.
From $M$, for each source we have calculated the source averaged H$_2$ volume and column densities,
$n_{\rm H_2}$ and $N_{t}({\rm H_2})$, assuming the clumps to be spherical and homogeneous. 
The H$_2$ column densities, $N_{p}({\rm H_2})$, have been estimated also from the 1.2~mm 
continuum peak flux using the equations in Beuther et al.~(\citeyear{beuther02}) and
adopting the same assumptions made to derive the mass (same $\beta$, dust mass 
opacity and gas-to-dust ratio). These represent average
values in the telescope beam of the 1.2~mm continuum observations ($\sim 24$\asec ), and we will use these 
estimates to derive the CO depletion in Sect.~\ref{depletion}, because the APEX \CII\ 
observations have a comparable angular resolution ($\sim 19$\asec ).
Also, we have computed the source averaged surface density, $\Sigma ({\rm H_2}) = 4 M/\pi (\theta_{\rm s} d)^2$, 
where $d$ and $\theta_{\rm s}$ are the source distance (see Table~\ref{sources}) and angular
diameter (see Table~\ref{tab_mass}), respectively. 
For 16435--4515c3, for which we did not have a kinematic source distance, we
computed it from the velocity of the \CII\ (3--2) line of the '--a' component (see Table~\ref{tab_lin_c18o}) 
following the method explained in Fontani et al.~(\citeyear{fonta05}). The method, which
assumes the rotation curve of Brand \& Blitz (\citeyear{brand}), is valid
for distances from the Galactic centre between 2 and 25 kpc, and provides two
kinematic distances, 3.1 and 12.6~kpc. As for the other sources with distance ambiguity,
we adopted the `near' value, i.e. 3.1~kpc.

The results are listed in Table~\ref{tab_mass}, where we give $M$, $n_{\rm H_2}$, 
$N({\rm H_2})$ (both $N_{t}({\rm H_2})$ and $N_{p}({\rm H_2})$)
and  $\Sigma ({\rm H_2})$. 
For completeness, in Cols.~2 and 3 of Table~\ref{tab_mass} we list
the full width half maximum diameters, $\theta_{\rm s}$ (deconvolved with the telescope beam  of 24\asec )
adopted to derive $n_{\rm H_2}$, $N_{t}({\rm H_2})$ and $\Sigma ({\rm H_2})$, as well as the integrated flux 
densities used to calculate the masses, $S_{\nu}$. Both parameters are taken from 
Beltr\'an et al.~(\citeyear{beltran06}).
The mean mass turns out to be 439 \solm , and the median is 244 \solm .
As expected, these values are systematically higher than those derived by  
Beltr\'an et al.~(\citeyear{beltran06}) given that the kinetic temperatures from 
ammonia (see Table~\ref{temperature}) are 
systematically lower than the representative value (30~K) assumed by
Beltr\'an et al.~(\citeyear{beltran06}). 
The clump-averaged column densities are of the order of $10^{22}-10^{23}$ \cmq , 
with a mean value of $1.6 \times 10^{23}$ \cmq\ and a median of $8.5 \times 10^{22}$ \cmq . 
The volumn densities are of the order of $10^{4}-10^{5}$ \cmc , with a mean value of 
$3.1 \times 10^{5}$ \cmc\ and a median of $3.9 \times 10^{4}$ \cmc .
Finally, the mean surface density is 0.37 g \cmq , and its median value is 0.19 g \cmq .
The difference between mean and median values for all parameters is due to the
clump 16573--4214c2, which has an angular diameter ($\sim 7$\arcsec )
more than twice smaller than any other source of the sample, despite
the comparable mass.

The clump-averaged column densities $N_{t}({\rm H_2})$ are generally smaller 
than the theoretical threshold given by Krumholz \& McKee~(2008) to avoid cloud 
fragmentation, which is 
$\sim 1$ g \cmq , corresponding to $\sim 3\times 10^{23}$ \cmq . 
Also, the distance independent parameter $\Sigma ({\rm H_2})$
is smaller, on average, than the theoretical values predicted for clumps that
are going to form high-mass stars or
super star clusters, which are expected to be larger than $\sim 0.7$ g \cmq\ 
(Chakrabarti \& McKee~\citeyear{cem}, Krumholz \& McKee~2008).
Rather, the values measured in this work are consistent to those
predicted for clumps that are going to form intermediate-mass stars 
and stellar clusters (Chakrabarti \& McKee~\citeyear{cem}).
We stress, however, that $\Sigma ({\rm H_2})$ and $N_{t}({\rm H_2})$
represent average values across the whole clumps, 
which could in reality be fragmented in smaller and denser cores.
Therefore, our $\Sigma ({\rm H_2})$ and $N_{t}({\rm H_2})$ are to be 
considered as lower limits for the individual embedded cores.
On the other hand, the column densities calculated from the 1.2~mm
continuum peak flux, $N_{p}({\rm H_2})$, are closer to the
theoretical threshold proposed by Krumholz \& McKee~(2008),
indicating that the central regions of the clumps could more easily form
massive stars.
Moreover, we have compared the clump masses to the threshold
proposed by Kauffmann \& Pillai~(\citeyear{kep}) based on an empirical 
mass-radius relation which predicts that a cloud must exceed the relation 
$M > M_{\rm thr} = 870 {\rm M_{\odot}} (R/{\rm pc})^{1.33}$ to form massive stars. The 
mass thresholds $M_{\rm thr}$ are listed in Col.~9 of Table~\ref{tab_mass}.
As $R$, we have used half of the diameters listed in the same Table.
We can see that our clump
masses are all above the predicted threshold, with four exceptions: 
13039--6108c6,  15557--5215c3, 16164--4929c3 and 16482--4443c2,
for which, in any case, the measured mass and the corresponding
threshold value are similar, and for 16482--4443c2 the mass threshold
is even an upper limit.
Based on these results, we claim that the targets have the potential to form 
massive stars.

\begin{table*}
\centering
\begin{minipage}{160mm}
\caption{Parameters derived from the 1.2~mm continuum emission: 1.2~mm continuum 
flux density ($S_{\nu}$) integrated over the 3$\sigma$ rms contour level, 
clump angular diameter ($\theta_{\rm s}$), gas mass ($M$),
H$_2$ volume ($n_{\rm H_2}$), column ($N({\rm H_2}$))
and mass surface density ($\Sigma({\rm H_2})$). 
The last column lists the mass threshold, $M_{\rm thr}$, for a cloud that can form massive stars
based on the empirical mass-radius relation proposed by Kauffmann \& Pillai~(\citeyear{kep},
see text).}
 \label{tab_mass}
\begin{tabular}{lllllllll}
\hline
       source &   $S_{\nu}$$^{a}$ &     $\theta_{\rm s}$$^{b}$  &   $M$ &  $n_{\rm H_2}$ & $N_{t}({\rm H_2})$$^{c}$ & $N_{p}({\rm H_2})$$^{d}$  & $\Sigma({\rm H_2})$  & $M_{\rm thr}$ \\
         & (Jy) & (\arcsec )  & (\solm ) & ($\times 10^{4}$ \cmc ) &  ($\times 10^{22}$ \cmq ) &  ($\times 10^{23}$ \cmq ) & (g \cmq ) & (\solm )  \\
\hline
       08477--4359c1  &       1.8     &      35.6   & 86.73 &  11.5  &   11.0  &  1.42   & 0.24 &    73 \\
       13039--6108c6 &        1.04   &       40.3 &   101.5 &	3.90 &  5.64 &  0.68     & 0.12 &        127 \\
       13560--6133c2  &      0.77    &       23.8   & 426.8 &  6.27  &  12.5  &  1.09        & 0.27 &   194 \\  
       14183--6050c3 &        1.06   &       39.6  &  207.7 &  2.96  &  5.95 &  0.68        & 0.13 &    196 \\  
    15038--5828c1 &        0.49   &      24.7  &   250.6 &  4.62   &   8.53  &  0.85        & 0.19 &      175 \\  
  15278--5620c2 &         1.5    &      40.1  &  243.7  & 3.34   &   6.81 &  0.82          & 0.15 &     200 \\  
  15470--5419c1     &    1.16        &    24.2     &  310.2 &   11.0   &   16.4 &  1.37    & 0.36 &     131 \\  
  15470--5419c3     &   3.06         &   54.1      &  743.4 &  2.37   &   7.84 &  1.11       & 0.17 &    382  \\  
  15557--5215c2     &    2.89        &    41.3     &  633.4 &  3.67   &   9.96 &  1.55      & 0.22 &    293 \\  
  15557--5215c3     &   0.51         &   35.8      &  194.3 &  1.73   &   4.07 &  0.49       & 0.09 &    242  \\  
  16061--5048c1     &    2.1          &   28.1      & 284.3  & 9.54   &   14.4  &  1.66      & 0.31 &   134 \\  
  16061--5048c4     &   1.57         &   62.8     &   504.2 &  1.52   &   5.12 &  1.22      & 0.11 &    391 \\  
  16093--5128c2     &   0.67         &   39.6     &   481.6 &  1.19   &   4.29  &  0.92     & 0.09 &    427 \\  
  16093--5128c8     &   0.86         &     40.1    &    642  & 1.52   &    5.57 &  0.68      & 0.12 &    434 \\  
       16164--4929c3  &     0.35    &    36.2  &     54.79 &  2.28   &   3.21 &  1.50       & 0.07 &      122 \\  
       16254--4844c1  &    1.15     &     20.3    &  164.2  & 17.4   &    17.9 &  1.43      & 0.39 &    81 \\  
      16435--4515c3   &   0.51      &    17.7    &   147 &   30.9 &   25.3 &  1.20   & 0.55 &    61 \\  
       16482--4443c2  & 0.24    &  $\ll 24$$^{e}$ & 59.08 & $\gg 3.81^{e}$ & $\gg 4.63^{e}$ &  0.66   & $\gg 0.10^{e}$ &  $\ll 101^{e}$ \\
       16573--4214c2  &     0.93    &     7.29  &    108.3	  & 553   &   157  &  1.89      & 3.4  & 14 \\  
       17040--3959c1  &     0.81    &    16.6    &    3428  & 5.80   &   23.7 &  1.60       & 0.52  & 505 \\ 
       17195--3811c2  &     2.61    &    44.3    &   597.9 &  5.12   &    12.2  &  1.77      & 0.27 &  246     \\  
 \hline
 \end{tabular}
 
$^{a}$  from Beltr\'an et al.~(\citeyear{beltran06}); \\
$^{b}$  full width half maximum angular diameters deconvolved with the telescope beam, from Beltr\'an et al.~(\citeyear{beltran06}); \\
$^{c}$  computed from the total clump mass (see text);\\
$^{d}$  computed from the 1.2~mm continuum peak flux (see text);\\
$^{e}$  unresolved in the SIMBA map. Therefore, $n_{\rm H_2}$, $N_{t}({\rm H_2})$ and $\Sigma({\rm H_2})$
are lower limits, and $M_{\rm thr}$ is an upper limit.  \\
 \end{minipage}
 \end{table*}

\subsection{\CII\ column densities and CO depletion factors}
\label{depletion}

The \CII\ column density has been derived from line intensities 
assuming optically thin lines and LTE conditions. 
Under these assumptions one can demonstrate that the
total column density of \CII , $N_{\rm C^{18}O}$, is related to the 
integrated intensity of a rotational transition $J \rightarrow J-1$ 
according to (e.g.~Pillai et al.~\citeyear{pillai07}):
\begin{eqnarray}
\lefteqn{N_{\rm C^{18}O}  =  \frac{N_J}{g_J}Q(T_{\rm ex}){\rm e}^{\left(\frac{E_J}{kT_{\rm ex}}\right)} = }  \nonumber \\
\lefteqn{  =  \frac{3 h}{8 \pi^3} 
  \frac{1}{S \mu^2} 
  \frac{I({\rm C^{18}O})}{J_{\nu}(T_{\rm ex})-J_{\nu}(T_{\rm BG})} 
  \frac{Q(T_{\rm ex}){\rm e}^{\left(\frac{E_J}{kT_{\rm ex}}\right)}}{{\rm e}^{h \nu / k T_{ex}}-1}
  \frac{1}{\eta_{\nu}} }
\label{eq_nj}
\end{eqnarray}
where: $I({\rm C^{18}O})$ is the integrated intensity of the line;
$N_J$ is the column density of the upper level;
$g_J$, $E_J$ and $S$ are statistical weight (=$2J+1$), 
energy of the upper level and line strength, respectively; $Q(T_{\rm ex})$ is 
the partition function at the temperature $T_{\rm ex}$; 
$\nu$ the line rest frequency; $J_{\nu}(T_{\rm ex})$ and $J_{\nu}(T_{\rm BG})$  
the equivalent Rayleigh-Jeans temperatures at frequency $\nu$
computed for the excitation and
background temperature ($T_{\rm BG}\sim 2.7~K$), respectively;
$\mu$ the molecule's dipole moment ($0.1102$ Debye for \CII );
$\eta_{\nu}$ the beam filling factor. 
The total integrated intensity $I({\rm C^{18}O})$ was derived from the
integral over all the channels with emission instead of the area given by the
Gaussian fits to take into account also non-Gaussian features. 

As for the derivation of the \H\ column density, 
we assumed a unity filling factor for all sources, so that 
$N_{\rm C^{18}O}$ is a beam-averaged value. 
As excitation temperature, we have used the kinetic temperatures listed in Col.~3 of
Table~\ref{temperature} (i.e. we are assuming LTE conditions for \CII\ (3--2)). 
The resulting column densities are listed in Table~\ref{tab_dep}.
To check if the assumption of optically thin lines is reasonable,
we have estimated the line optical depths from the solution of the line
radiative transfer equation:
$$
\tau=-{\rm ln}\left(1-\frac{T_{\rm MB}}{J_{\nu}(T_{\rm ex})-J_{\nu}(T_{\rm BG})}\right)\;
$$
and found values smaller than $\simeq 0.6$, consistent with our
assumption of optically thin lines.

The CO depletion factor, $f_{\rm D}$, is
defined as the ratio between the 'expected' abundance of CO relative to 
H$_2$ ($X_{\rm C^{18}O}^{E}$) and the 'observed' value:
\begin{equation}
f_{\rm D}=\frac{X_{\rm C^{18}O}^{E}}{X_{\rm C^{18}O}^{O}}\;\;.
\end{equation}
$X_{\rm C^{18}O}^{O}$ is the ratio between the observed \CII\ 
column density and the observed H$_2$ column density. 
For this latter, we have used the value derived from the 1.2~mm
continuum peak flux, $N_{p}({\rm H_2})$ (Col.~7 of Table ~\ref{tab_mass}), 
which is averaged over a beam comparable to that of the \CII\
observations (24\asec\ against 19\asec ).

To compute $X_{\rm C^{18}O}^{E}$, we have 
taken into account the variation of atomic carbon and oxygen abundances 
with distance to the Galactic Center following the same
procedure adopted in Fontani et al.~(\citeyear{fonta06}; see 
also Miettinen et al.~\citeyear{miettinen}). 
Assuming the canonical abundance of $\sim 8.5 \times 10^{-5}$ 
for the abundance of the main CO isotopologue in the 
neighbourhood of the solar system (Frerking et 
al.~\citeyear{frerking}, see also Langer et al.~\citeyear{langer}
and Pineda et al.~\citeyear{pineda}), we have computed the expected 
CO abundance at the Galactocentric distance ($D_{\rm GC}$) of
each source according to the relationship:
\begin{equation}
X_{\rm C^{16}O}^{E}=8.5\times 10^{-5}{\rm exp}\,(1.105-0.13\,D_{\rm GC}({\rm kpc}))\;,
\end{equation}
which has been derived according to the abundance gradients
in the Galactic Disk for $^{12}$C/H and $^{16}$O/H listed in Table~1 
of Wilson \& Matteucci~(\citeyear{wem}), and assuming that the Sun has a
distance of 8.5 kpc from the Galactic Centre. Then, following 
Wilson \& Rood~(\citeyear{wer}), we have assumed that the Oxygen
isotope ratio $^{16}$O/$^{18}$O depends on $D_{\rm GC}$ 
according to the relationship $^{16}$O/$^{18}$O=$58.8\times D_{\rm GC}({\rm kpc})+37.1$,
which gives:
\begin{equation}
X_{\rm C^{18}O}^{E}=\frac{X_{\rm C^{16}O}^{E}}{(58.8\,D_{\rm GC}+37.1)}\;\;\;.
\label{eq_abb}
\end{equation}

The results are listed in Table~\ref{tab_dep}: 
Cols.~2, 3 and 4 list the \CII\ (3--2) integrated line intensity ($I({\rm C^{18}O})$), 
the \CII\ column density ($N({\rm C^{18}O})$) and
the observed \CII\ abundance ($X_{\rm C^{18}O}^{O}$); Cols.~5 and 6
give the expected \CII\ abundance ($X_{\rm C^{18}O}^{E}$) calculated
at the Galactocentric distance of the source (Table~\ref{sources}) and the
CO depletion factor ($f_{\rm D}=X_{\rm C^{18}O}^{E}/X_{\rm C^{18}O}^{O}$),
respectively. 

The mean $f_{\rm D}$ is $\sim 32$ and the median value is 29.
These values are remarkably higher than those obtained
in other IRDCs (Pillai et al.~\citeyear{pillai07}, Miettinen
et al.~\citeyear{miettinen}, Hern\'andez et al.~\citeyear{hernandez}), 
and are among the highest obtained
both in low-mass starless cores (Crapsi et al.~\citeyear{crapsi},
Tafalla et al.~\citeyear{tafalla06}) and in
massive clumps from observations with low-angular resolution
(Fontani et al.~\citeyear{fonta06}, Chen et al.~\citeyear{chen}).

Such a difference could be explained by the fact that the
transition used in this work to derive $f_{\rm D}$ has a critical
density of $\sim 5 \times 10^{4}$ \cmc , comparable to that
of the central region where the CO freeze-out is expected to be
important. In fact, the CO freeze-out timescale, which depends
on the H$_2$ volume density, becomes shorter
than the free-fall timescale at densities larger than
a few $10^{4}$ \cmc\ (see e.g. Bergin \& Tafalla~\citeyear{bet07},
Caselli~\citeyear{caselli11}). On the contrary, in most
of the studies performed so far $f_{\rm D}$ was derived from
the CO (1--0) or (2--1) lines, which trace lower-density
gas where CO is significantly non-depleted, resulting in
values of $f_{\rm D}$ smaller than or comparable to $\sim 10$ 
(e.g.~Crapsi et al.~\citeyear{crapsi},
Tafalla et al.~\citeyear{tafalla06}, Pillai et al.~\citeyear{pillai07}). 
However, the discussion of this result requires three main comments.
First, the values of the 'canonical' CO abundance measured by other 
authors in different objects vary by a factor of 2 (see e.g. Lacy et al.~1994; 
Alves et al.~1999), but because the value assumed by us was
derived from the Taurus star forming regions (Frerking et al.~\citeyear{frerking})
and used in previous estimates (Crapsi et al.~\citeyear{crapsi};
Emprechtinger et al.~\citeyear{emprechtinger}),
it is likely the most appropriate to make comparison with low-mass dense
cores. 
Second, the angular resolution of our observations 
allows us to derive only average values of $f_{\rm D}$ over angular
regions much larger than the typical fragmentation scales (few arcseconds
at the distance of our targets), which in reality may have complex structures,
so that our estimates should be considered as lower limits. 
Finally, high depletion of \CII\ could be due to mechanisms
other than freeze-out in regions where young stellar
objects are embedded, and can decrease the CO abundance even
in warm environments (for a detailed explanation, see
Fuente et al.~\citeyear{fuente}). This comment especially concerns
the clumps where the average kinetic temperature exceeds
the CO sublimation temperature of $\sim 20$~K.

\begin{table*}
\centering
\begin{minipage}{140mm}
\caption{Parameters used to determine the CO depletion factor: integrated 
intensity of the \CII\ (3--2) line ($I({\rm C^{18}O}))$), \CII\ total column density ($N({\rm C^{18}O})$),
observed \CII\ abundance ($X_{\rm C^{18}O}^{O}$), 'expected' 
\CII\ abundance ($X_{\rm C^{18}O}^{E}$), and CO depletion factor 
($f_{\rm D}$ = $X_{\rm C^{18}O}^{E}$/$X_{\rm C^{18}O}^{O}$).
The errors are given between parentheses. The uncertainty on $f_{\rm D}$
based on the calibration errors affecting the \CII\ and H$_2$ column densities is of the
order of the 40-50$\%$.}
\label{tab_dep}
\begin{tabular}{llllll}
\hline
source &  $I({\rm C^{18}O})$  & $N({\rm C^{18}O})$ & $X_{\rm C^{18}O}^{O}$ & $X_{\rm C^{18}O}^{E}$ & $f_{\rm D}$ \\
       &  (K \kms )  &  ($\times 10^{15}$\cmq ) & ($\times 10^{-8}$) & ($\times 10^{-7}$) &  \\
\hline
08477--4359c1  & 6.5(0.1) & 3.03(0.05) &    2.14 &	         1.45 &   7 \\ 
13039--6108c6  & 1.5(0.2)  & 0.65(0.09)  &  0.95 &	         2.08 &   22 \\
13560--6133c2  & 1.8(0.1)  & 1.08(0.06) &   0.99 &	         2.68 &   27 \\
14183--6050c3  & 0.9(0.1)  & 0.41(0.04) &   0.60 &	         2.54 &   42 \\
15038--5828c1  & 0.89(0.09) & 0.52(0.05) &  0.61 &	         3.30 &   54  \\
15278--5620c2  & 2.5(0.1)  & 1.10(0.04) &   1.34 &	         2.93 &   22 \\
15470--5419c1  & 2.3(0.1)  & 1.36(0.06)  &  0.99 &	        3.46 &    35 \\
15470--5419c3  & 2.6(0.1)  & 1.06(0.04) &   0.96 &	        3.46 &   36 \\			  
15557--5215c2  & 4.2(0.1)  & 1.90(0.04) &   1.20 &	        3.79 &   32  \\
15557--5215c3  & 1.6(0.1)  & 0.76(0.05) &   1.57 &	        3.79 &   24  \\
16061--5048c1  & 8.8(0.2)  & 4.7(0.1) &     2.83 &	        3.39 &   12  \\
16061--5048c4  & 3.0(0.1)  & 1.23(0.04) &   1.01 &	        3.39 &   34  \\			       
16093--5128c2--a & 3.55(0.08) & 1.59(0.04) &  1.72 &     5.09 &    29  \\
16093--5128c8  & 1.00(0.06)  & 0.44(0.03) &  0.65    &	5.09 &   78  \\
16164--4929c3  & 3.01(0.07)  & 1.40(0.03) &  0.93  &	2.80 &   30  \\			       
16254--4844c1  & 2.82(0.09)  & 1.93(0.06) &  1.35  &	3.39 &   25  \\
16435--4515c3--a & 1.73(0.09) & 0.68(0.07) &  0.57 &	4.13 &   73  \\
16482--4443c2  & 4.7(0.1)  & 2.96(0.06) &   4.45 &   3.87 &    9  \\
16573--4214c2  & 7.8(0.1)  & 2.2(0.3)  &   1.18 &	         2.98 &   25   \\
17040--3959c1  & 6.32(0.09)  & 5.32(0.08) &  3.32  &	1.59 &    5  \\
17195--3811c2  & 3.07(0.08)  &  1.33(0.03) &  0.75 &	4.04 &  54 \\
\hline
\end{tabular}

\end{minipage}
\end{table*}

\section{Discussion}
\label{discu}

\subsection{Virial state of the clumps}
\label{dis_virial}

In order to investigate if our targets are gravitationally stable, we have computed the
virial masses of the clumps assuming virial equilibrium and negligible
contributions of magnetic field and surface pressure. 
Assuming also that the cores are spherical, 
one can demonstrate that the gas mass is given by:

\begin{equation}
M_{\rm VIR}(M_{\odot})\simeq k\, \Delta v^{2} {\rm (\kms )}\,R ({\rm pc})\;\;, 
\end{equation}
where $R$ is the clump radius, $\Delta v^{2}$ the line width and $k$ is a 
multiplicative factor that depends on the gas density distribution as a function 
of the distance to the clump centre (see Eq. (3) in MacLaren et al.~\citeyear{maclaren}). 
For a homogeneous clump (i.e. $\rho = const.$),
$k\simeq 210$, while for $\rho \sim r^{-2}$,  $k\simeq 126$.
The millimetre continuum maps do not allow to derive the
density profile of the clumps because of the low angular resolution
and sensitivity to extended emission. However, previous studies suggest that
the density structure of massive clumps reasonably can be approximated
by a constant density in the inner regions, and by a steep
power-law in the outer layers (see e.g.~Beuther et al.~\citeyear{beuther02}, 
Fontani et al.~\citeyear{fontani02},
Hill et al.~\citeyear{hill05}).

\begin{figure}
\begin{minipage}{90mm}
 \begin{center}
 \resizebox{\hsize}{!}{\includegraphics[angle=0]{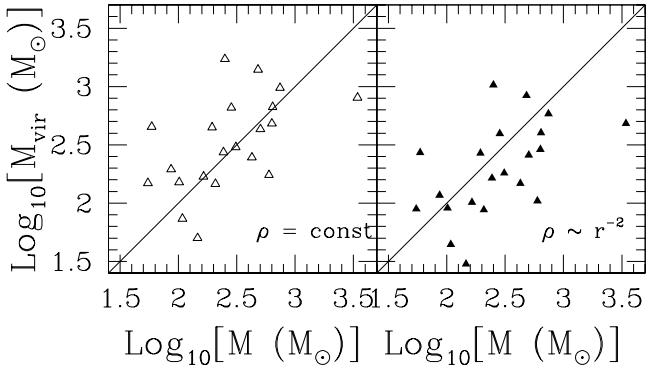}}
 \caption[]
 {\label{mvir_m}{Virial masses against gas masses computed from the 1.2~mm
 dust continuum emission. \mvir\ is calculated assuming a density distribution
 of the type $\rho = const.$ (left panel) and $\rho \sim r^{-2}$
 (right panel). In both panels, the line indicates \mvir = $M$.}}
 \end{center}
 \end{minipage}
\end{figure}

\begin{figure}
\begin{minipage}{90mm}
 \begin{center}
 \resizebox{\hsize}{!}{\includegraphics[angle=0]{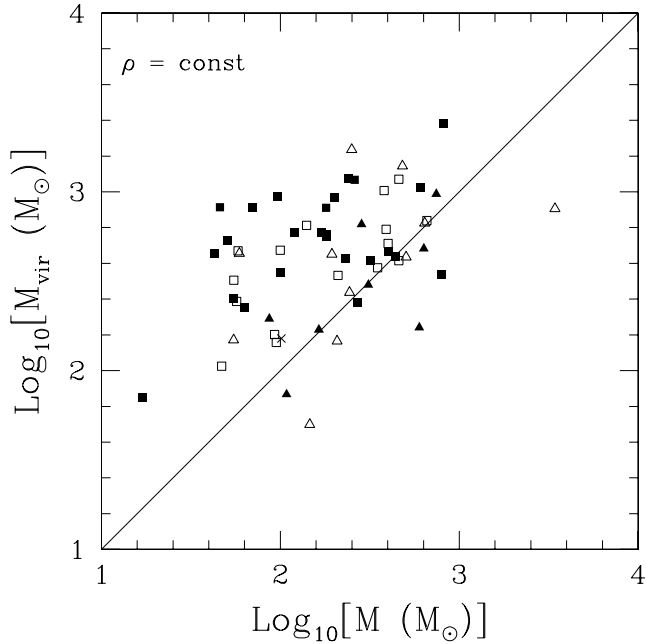}}
 \caption[]
 {\label{mvir_m_sepulcre}{Virial masses against gas masses computed from the dust continuum
 emission for the clumps studied in this work (triangles) and the massive
 clumps observed by L\'opez-Sepulcre et al.~(\citeyear{sepulcre}, squares).
 Filled and open triangles represent sources detected and undetected at
 24 $\mu$m, respectively, while the filled and open squares represent
 the infrared-bright and infrared-dark sources selected by L\'opez-Sepulcre et al.~(\citeyear{sepulcre}),
 respectively. The cross indicates
 the clump with no 24 $\mu$m image available, 13039--6108c6.
 \mvir\ is calculated assuming homogeneous clumps. The line indicates \mvir = $M$.}}
 \end{center}
 \end{minipage}
\end{figure}

In Fig.~\ref{mvir_m} we compare the mass derived from the dust continuum
emission to \mvir\ obtained assuming $\rho \sim const.$ (left panel)
and $\rho\sim r^{-2}$ (right panel). 
We note that \mvir\ is on average larger than $M$ for the case $\rho \sim const.$, 
while it is on average smaller for the other case. Because the overall density
distribution of the clumps is likely in between these two extreme
cases, we suggest that, on average, the clumps are close to virial equilibium. 
Our findings are in good agreement with those
derived by L\'opez-Sepulcre et al.~(\citeyear{sepulcre}) in a sample
of high-mass clumps supposed to span a wide range of evolutionary stages.
The gas masses of the high-mass clumps studied by
L\'opez-Sepulcre et al.~(\citeyear{sepulcre}) were derived from 
millimetre continuum measurements, and the 
dust temperatures assumed are in agreement with ours (see their Table~1).
Moreover, they estimated the virial masses from \CII\ (2--1) line widths and
assuming $\rho \sim const.$
As shown in Fig.~\ref{mvir_m_sepulcre}, we do not highlight systematic differences
between the two sub-samples, and between infrared-dark and infrared-bright
sources either. On average, the sources of the L\'opez-Sepulcre et al.'s sample
appear to have slightly larger virial masses, so that potentially they might be 
less gravitationally bound than the clumps studied in this work.
However, we stress that the parameters from which the mass estimates 
are derived are affected by large errors
(especially the dust mass opacity, the distance and the gas-to-dust ratio)
and difficult to quantify. For example, the dust mass opacity coefficient
can introduce a factor 2 in the uncertainty (Beuther et al.~\citeyear{beuther02}).
Hence our conclusions on the virial stability of the targets must be taken 
with caution. 

\subsection{24 $\mu$m dark versus 24 $\mu$m bright clumps}
\label{sec_histo}

\begin{figure*}
\begin{minipage}{160mm}
 \begin{center}
 \resizebox{\hsize}{!}{\includegraphics[angle=0]{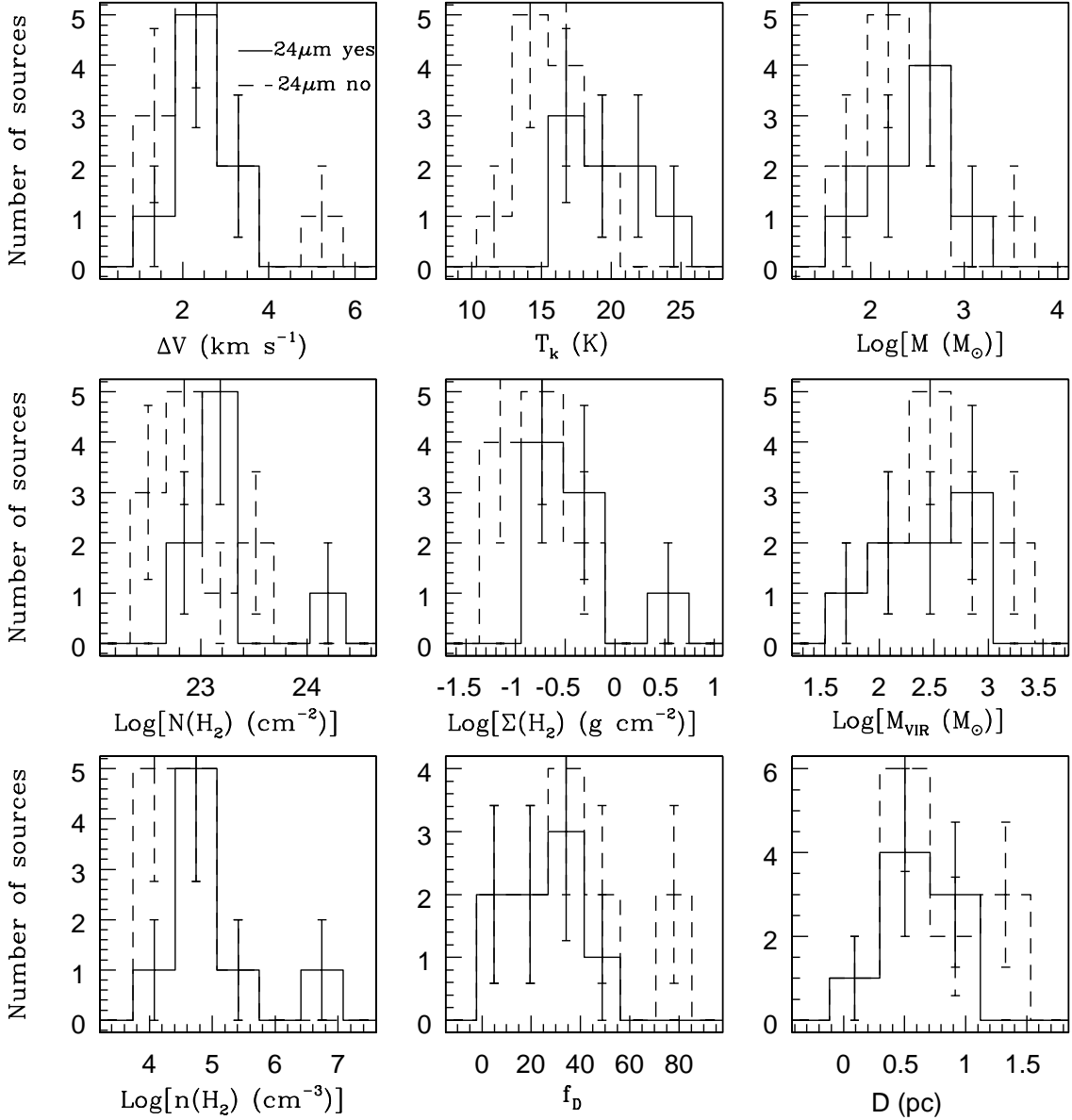}}
 \caption[]
 {\label{fig_histo}{Histograms comparing some of the physical and
 chemical properties of the clumps detected
 (solid line) and undetected (dashed line) in the Spitzer-MIPS 24 $\mu$m
 image (see Figs.~\ref{appA-fig1}). In the panels
 showing $N$(H$_2$), $n$(H$_2$) and $\Sigma$(H$_2$) we
 have not included the lower limits calculated for 16482--4443c2.}}
 \end{center}
 \end{minipage}
\end{figure*}

The presence of mid-infrared emission in molecular clumps is
very often the sign of embedded star formation activity. Therefore,
the IRDCs detected at 24 $\mu$m could harbour objects
more evolved than those undetected. If this is the case, 
the two groups should have physical 
properties indicative of a different evolutionary stage.
In Fig.~\ref{fig_histo} we show histograms which compare some
physical and chemical properties of the IRDCs detected and
undetected at 24 $\mu$m: line widths,
kinetic temperature, column ($N_{t}$(H$_2$)), volume and surface density of H$_2$, 
gas masses (both from continuum and the virial theorem), CO depletion factor
and clump diameter. For \mvir , we have considered the values calculated
assuming $\rho \sim const$, bearing in mind that those obtained
in the case $\rho \sim r^{-2}$ are just systematically lower by a factor
$\sim 1.7$. For the unresolved source 16482--4443c2,
we have not included the $n$(H$_2$) and 
$\Sigma$(H$_2$) lower limits, and replaced $N_{t}$(H$_2$) with $N_{p}$(H$_2$).
Although the statistics is poor because the two
sub-samples contain eight 24 $\mu$m bright and twelve
24 $\mu$m dark objects, the comparative histograms are useful to check if
one can notice clear systematic differences.
An inspection of the comparative histograms indicates that
the 24 $\mu$m dark clumps have lower \Tk\ and higher $f_{\rm D}$.
In fact, the average \Tk\ and $f_{\rm D}$ of the infrared-bright group turns out
to be $\sim 20$ K and 28, respectively, while that of the infrared-dark group is 
$\sim 16$ K and 35, respectively.
Also, if we exclude from the statistical analysis the clear outliers
17040--3959c1 and 16573--4214c2, the 24 $\mu$m dark clumps tend to have
lower mass (average $M$ of $\sim 260$ \solm\ from the 1.2~mm continuum), 
smaller source-averaged H$_2$ column density 
(average $N_{t}$(H$_2$) of $8.9\times 10^{22}$ \cmq ),
and smaller H$_2$ surface density (average $\Sigma$(H$_2$) of 0.19 gr \cmq ) than
the infrared-bright clumps (average $M$ of $\sim 370$ \solm , average 
 $N_{t}$(H$_2$) of $1.3\times 10^{23}$ \cmq , average $\Sigma$(H$_2$) of 0.27 g \cmq ). 

The fact that on average the group of the 24 $\mu$m-dark sources has
lower \Tk\ and higher CO depletion factor than the 24 $\mu$m-bright objects 
suggests that it likely contains objects less evolved, and this is in accordance with other
comparative studies of massive clumps with or without
indications of embedded star formation (Hill et al.~\citeyear{hill05}). 
However, the dark sources also have lower $M$ and H$_2$ column and surface 
density, and this could indicate that the 24 $\mu$m dark clumps may be destined 
to form less massive stars/clusters, thus explaining the non-detection at mid-infrared
wavelengths, although material could still be accreting.

\subsection{Relation between \Tk\ and line widths}
\label{dis_t_dv}

\begin{figure}
\begin{minipage}{90mm}
 \begin{center}
 \centerline{\includegraphics[angle=0,width=6.0cm]{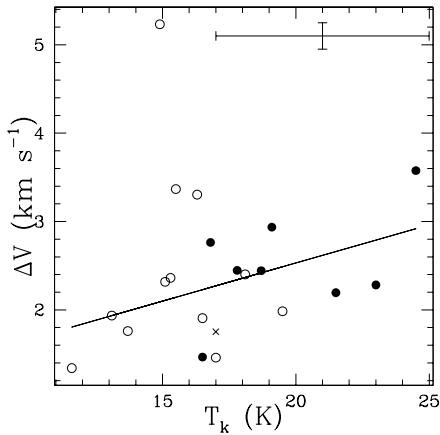}}
 \caption[]
 {\label{T_dv}{Gas kinetic temperature, \Tk , derived from ammonia, against
 the \CII\ line widths, $\Delta V$. The straight line is a least square fit to the data
 excluding 15038--5828c1, the clump with $\Delta V > 5$ \kms , i.e. much higher than
 any other clump and based on a very noisy spectrum (see Fig.~\ref{spectra_co_only}). 
 The slope of the linear fit is $\sim 0.09$. Filled and open symbols represent
 clumps detected and undetected at 24 $\mu$m, respectively. The cross indicates
 the clump with no 24 $\mu$m image available, 13039--6108c6. Typical errorbars
 are depicted in the top-right corner.}}
 \end{center}
 \end{minipage}
\end{figure}

Two indicators of active star formation are the line width and
the gas temperature, which are both expected to become higher
with increasing star formation activity.
In fact, the line widths of dense gas tracers are found to be 
higher in more evolved star-forming clumps than in quiescent 
regions of infrared-dark clouds (e.g.~Hill et al.~\citeyear{hill10},
Ragan et al.~\citeyear{ragan12}),
while the cores associated with protostars are on average warmer
than the starless cores, both in low- and high-mass star forming regions 
(e.g. Foster et al.~\citeyear{foster}, Emprechtinger et al.~\citeyear{emprechtinger},
Ragan et al.~\citeyear{ragan11}).
Fig.~\ref{T_dv} shows that there is a slight correlation between \Tk\ and
the \CII\ line width. If we exclude from the statistical analysis
15038--5828c1, for which $\Delta V$ is much higher than that
of the other targets and has a spectrum with poor S/N (see Fig.~\ref{spectra_co_only}),
a linear least square fit to the data yields a slope of $\sim 0.09$.
Statistical correlation between \Tk\ and line width can be investigated
also through non-parametric statistical methods, like the Kendall's $\tau$ correlation
coefficient. This measures the rank correlation, namely how two quantities are ordered 
similarly when ranked by each of them\footnote{for details see, e.g.,~http://www.statsoft.com/textbook/nonparametric-statistics/)}. 
$\tau$ can range between 1 (perfect agreement between 
the two rankings) and $-1$ (one ranking is the reverse of the other).
We find $\tau\sim 0.23$ between \Tk\ and line width. 
These results suggest a faint correlation
between the two parameters, which indicates that the warmer clumps tend
to also have larger line widths, i.e. tend to be more turbulent.
However, the large dispersion cannot allow us to draw any firm conclusion.

\subsection{Relation between CO depletion and other physical parameters}
\label{dis_dep}

One of the main results of this work is the high
CO depletion factor measured in our targets, with values
higher than those found in comparable high-mass clumps (e.g.
Miettinen et al.~\citeyear{miettinen}, Chen et al.~\citeyear{chen}), and even in
low-mass pre--stellar cores (see e.g. Crapsi et al.~\citeyear{crapsi}).
The fact that we have derived $f_{\rm D}$ from the \CII\ (3--2)
transition, which has a critical density of $5 \times 10^{4}$ \cmc , 
certainly makes the difference with respect to previous works in which
$f_{\rm D}$ was calculated from the lower excitation transitions tracing
material where freeze-out of CO is expected to be less important.
In fact, freeze-out of CO and other neutrals onto dust grains is favoured 
in gas characterised by low temperatures ($T\leq 20~K$)
and high densities ($n \geq 10^{5}$ \cmc ), in which the freeze-out
timescale is generally shorter than the free-fall timescale 
(e.g.~Bergin \& Tafalla~\citeyear{bet07}). In fact, high CO depletion 
factors (of the order of 10 or more) were measured towards the dense and
cold nuclei of low-mass pre--stellar cores
(see e.g. Crapsi et al.~\citeyear{crapsi}).
After the formation of the protostellar object(s) at the core centre, evaporation of 
CO starts and the CO depletion factor drops (Caselli et al.~\citeyear{caselli02b}).
This theoretical prediction has been partly confirmed
by observations of both low- and high-mass dense cores 
(e.g. Emprechtinger et al.~\citeyear{emprechtinger}, 
Fontani et al.~\citeyear{fonta06}).

\begin{figure*}
\begin{minipage}{160mm}
\begin{center}
\centerline{\includegraphics[angle=0,width=8.0cm]{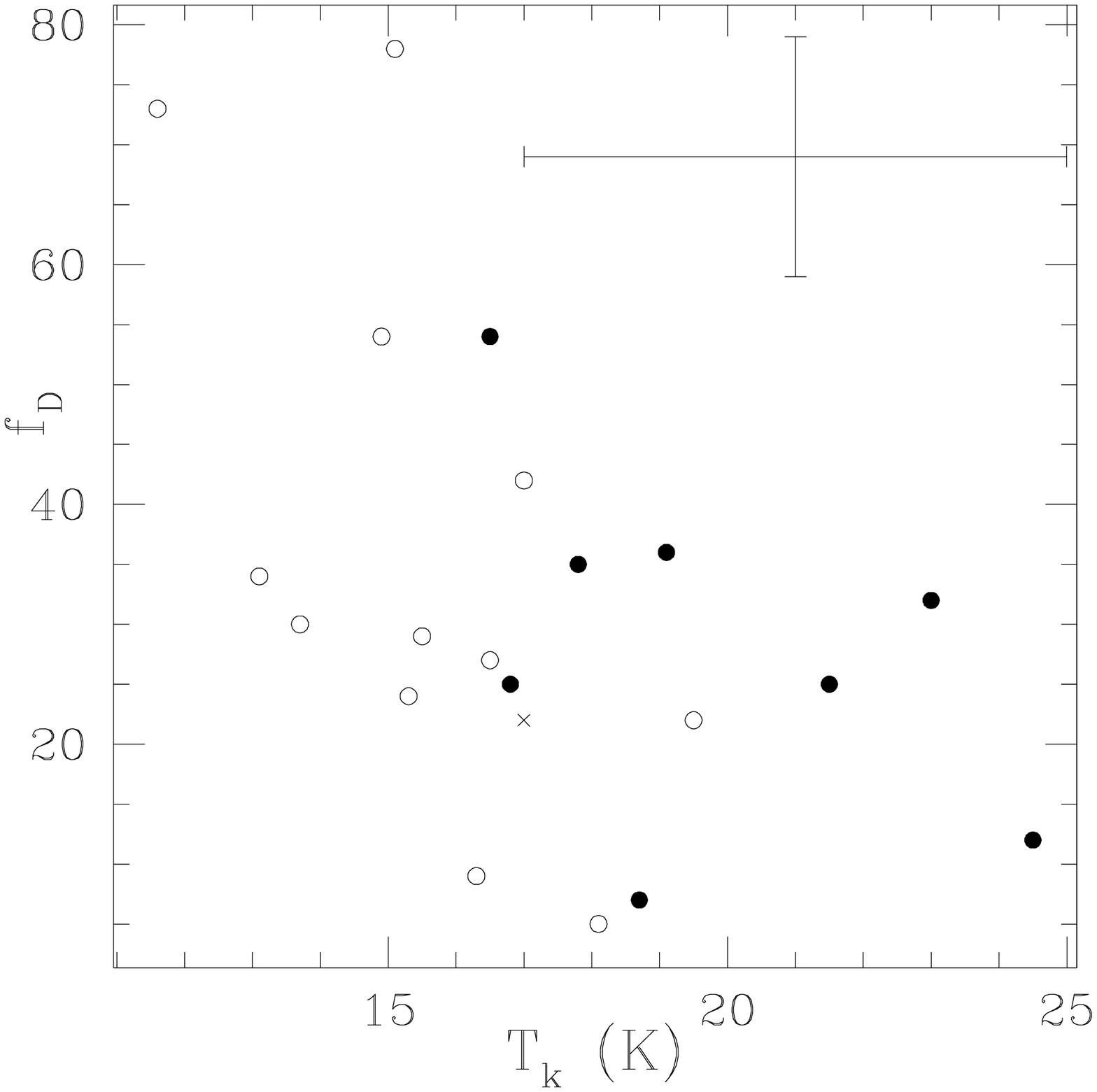}
		 \includegraphics[angle=0,width=8.0cm]{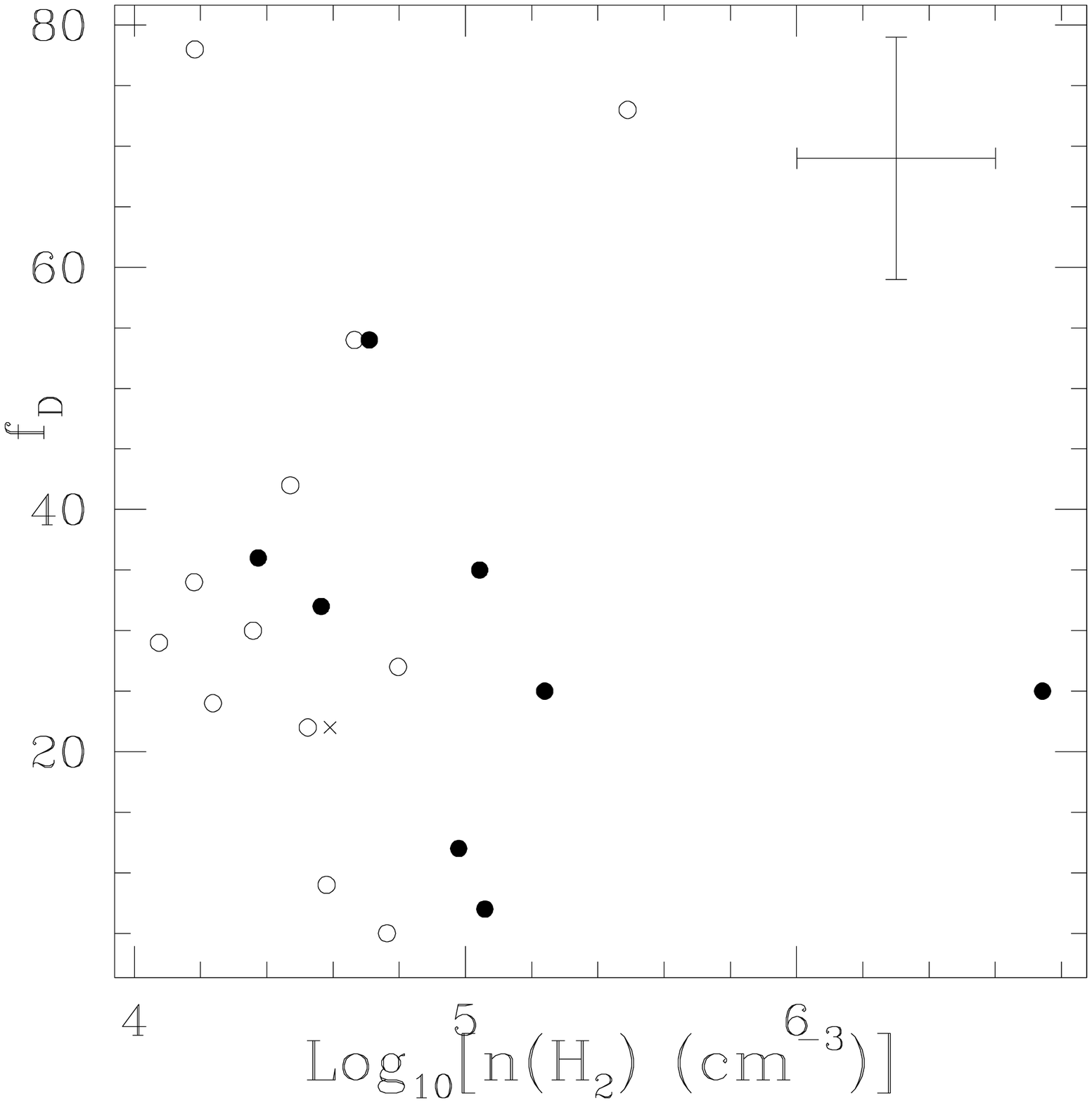}}
 \caption[]
 {\label{dep_t_n}{CO depletion factor ($f_{\rm D}$) against gas kinetic
 temperature (left panel) and H$_2$ volume density (right panel). 
 The symbols have the same meaning as in Fig.~\ref{T_dv}.}}
 \end{center}
 \end{minipage}
\end{figure*}

In Fig.~\ref{dep_t_n} we show
$f_{\rm D}$ against \Tk\ (in the left panel) and $n{\rm (H_2)}$
(in the right panel). At a first glance, the two plots do not reveal
clear (anti-)correlations between the parameters. 
As made in Sect.~\ref{dis_t_dv}, we have performed a closer 
inspection of the data using the non-parametric ranking statistical 
test Kendall's $\tau$ (see Sect.~\ref{dis_t_dv}). 
If we consider all points, we find a faint anti-correlation between
$f_{\rm D}$ and \Tk\ (Kendall's $\tau$ = --0.34)
and also between $f_{\rm D}$ and $n{\rm (H_2)}$
(Kendall's $\tau$ = --0.12). The anti-correlation between
\Tk\ and $f_{\rm D}$ is consistent with the increase of CO depletion with
decreasing temperature, as already found in both low- and high-mass
dense cores (Emprechtinger et al.~\citeyear{emprechtinger}, 
Fontani et al.~\citeyear{fonta06}), while that between
$f_{\rm D}$ and $n{\rm (H_2)}$ is the opposite of what
expected from chemical models.
Even if we exclude from the statistical analysis 16573--4214c2, 
which has $n{\rm (H_2)}$ markedly much higher than that of the 
other members of the sample, the correlation between
$f_{\rm D}$ and $n{\rm (H_2)}$
is not observed (Kendall's $\tau$ = --0.11).
We point out, however, that all the parameters obtained
are average values over angular regions much larger than the
size expected for the embedded condensations, while the 
correlations are predicted for single cores.
Moreover, clumps embedded in different clouds may be
affected by different environmental conditions, so increasing
the dispersion. In fact, Emprechtinger et al.~(\citeyear{emprechtinger})
found that dense cores in Perseus showed the best trends between
the physical parameters, whereas when including cores from other 
star-forming regions the dispersion was significantly larger.

\begin{figure}
\begin{minipage}{70mm}
{\includegraphics[angle=0,width=7.0cm]{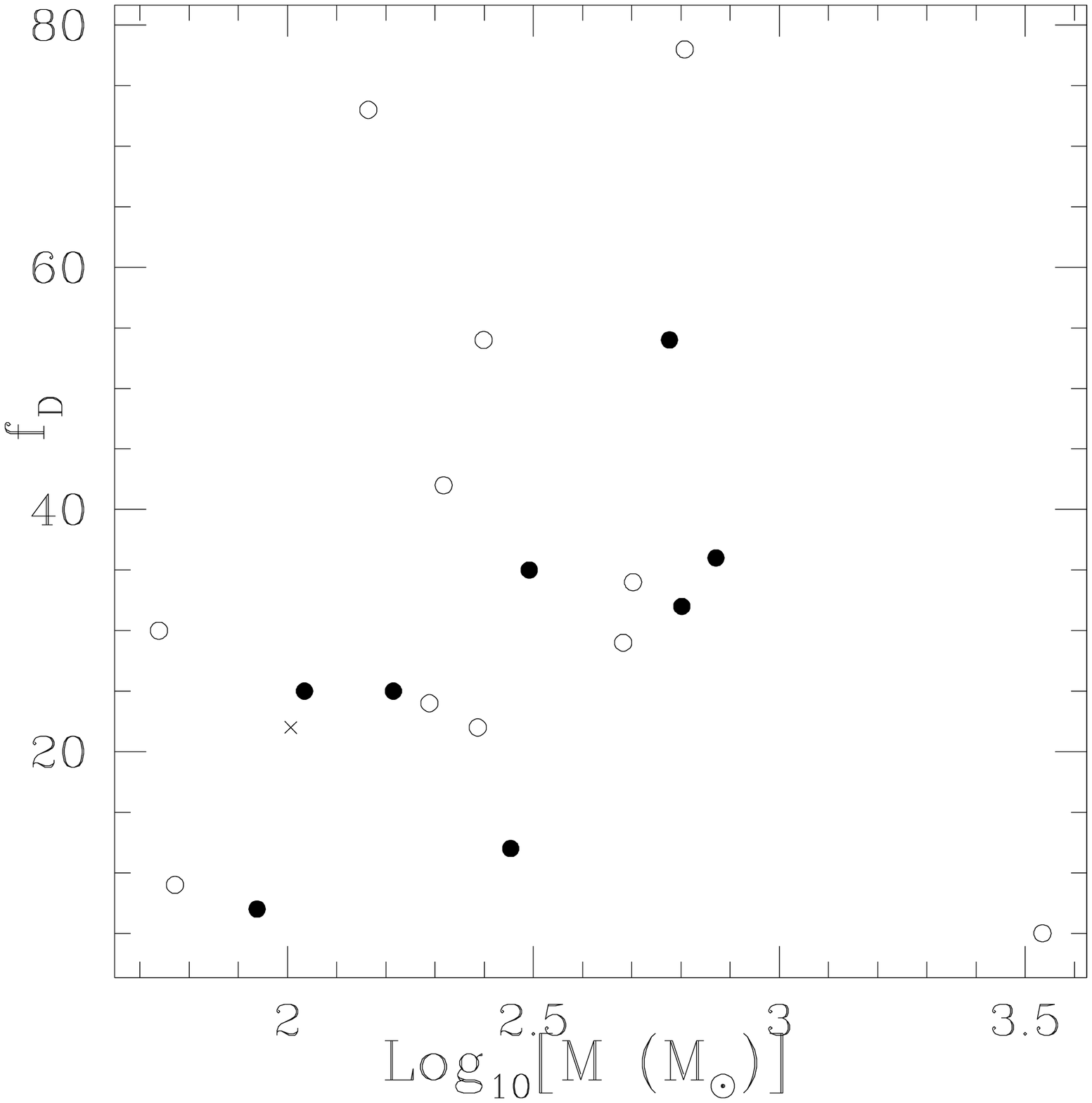}}
{\includegraphics[angle=0,width=7.0cm]{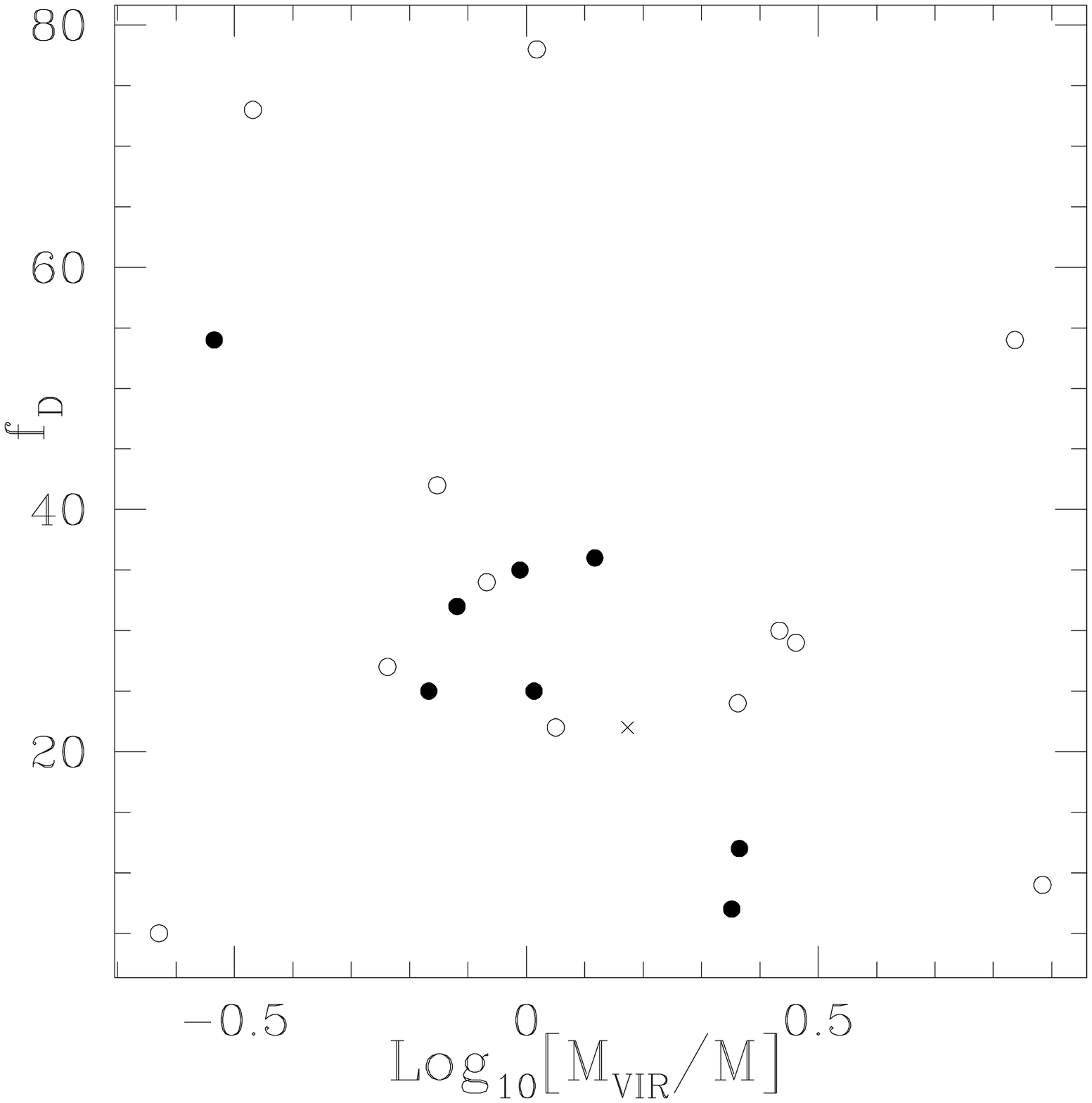}}
 \caption[]
 {\label{dep_par}{CO depletion factor ($f_{\rm D}$) against gas 
 mass derived from the 1.2~mm continuum (top panel) and \mvir /$M$ mass ratio
 (bottom panel). The symbols have the same meaning as in Fig.~\ref{T_dv}.}}
 \end{minipage}
\end{figure}

Finally, Fig.~\ref{dep_par} suggests that $f_{\rm D}$ is not clearly (anti-)correlated 
to either the clump mass $M$ and and the ratio \mvir /$M$ which
should show the tendency of a clump to be stable against gravitational
collapse. When excluding the clear
outlier 17040--3959c1 (whose mass is statistically much larger than
the rest of the sample), a slight correlation is found between $f_{\rm D}$
and $M$ (Kendall's $\tau$ = 0.4). However, again the large dispersion
of the data and the big errors on both parameters do not allow us to claim 
firm conclusions.

\section{Conclusions}
\label{conc}

We have undertaken a molecular line study, through observations of
\CII\ (3--2), \AMM\ (1,1) and (2,2),
\H\ (3--2) and \D\ (4--3), of 21 IRDCs in the southern hemisphere with the aim of
characterising the initial conditions of the high-mass star and cluster formation process. 
The sample targeted was selected from the high-mass millimetre clumps not
detected in the MSX images identified by Beltr\'an et al.~(2006), and includes sources 
with and without Spitzer 24 $\mu$m emission. The \AMM\ inversion transitions 
have been observed with the ATCA. The rotational transitions of the other molecular 
species have been observed with the APEX Telescope. We have detected \CII\ and
ammonia emission in all clumps, and \H\ emission in all the 12 sources observed
in this line. Only one source has been marginally detected in \D\ (4--3), which
is, to our knowledge, the first detection of this line in an IRDC.
The clumps have a median mass of $\sim 244$ \solm , appear to be
gravitationally bound and possess mass,
H$_2$ column and surface densities consistent with being potentially
the birthplace of high-mass stars.
The most striking result of the work is the high average CO depletion
factor (derived from the expected \CII -to-H$_2$ column density ratio
 compared to the observed value), which is
in between 5 and 78, with a mean value of 32 and a median of 29. 
These values, derived from the \CII\ (3--2) transition (which traces gas
denser than the lower-excitation lines commonly used in previous works), 
are larger than the typical CO depletion factors measured towards other 
IRDCs, and are comparable to or larger than the values derived in the low-mass 
pre--stellar cores closest to the onset of gravitational collapse.
Also, a faint anti-correlation is found between $f_{\rm D}$ and the gas
kinetic temperature.
These results suggest that the earliest phases of the high-mass star and stellar
cluster formation process are characterised by CO
depletions larger than in low-mass pre--stellar cores.
The clumps have an average temperature around 17~K.
We have found marginal statistical differences between the clumps
detected and undetected in the 24 $\mu$m Spitzer images, with
the Spitzer-dark clumps being on average colder, less massive and having 
lower H$_2$ column and surface densities, but higher CO depletion factors.
This indicates that the Spitzer-dark clumps are either
less evolved or destined to
form stars and stellar clusters less massive than the Spitzer-bright ones.

\section*{Acknowledgments}

This publication is based on data acquired with 
the Atacama Pathfinder Experiment (APEX). The Atacama Pathfinder 
Experiment is a collaboration between the Max-Planck-Institut f\"ur Radioastronomie, 
the European Southern Observatory, and the Onsala Space Observatory.
We thank the staff at the APEX telescope for performing the service mode observations 
presented in this paper. FF is deeply grateful to Ana L\'opez-Sepulcre for providing
us with the parameters of the massive cores in L\'opez-Sepulcre et al.~(\citeyear{sepulcre}).

{}

\newpage

\appendix

\section[]{Spitzer and SIMBA images}
\label{appendixA}

\begin{figure*}
\begin{minipage}{160mm}
 \begin{center}
 \resizebox{\hsize}{!}{\includegraphics[angle=-90]{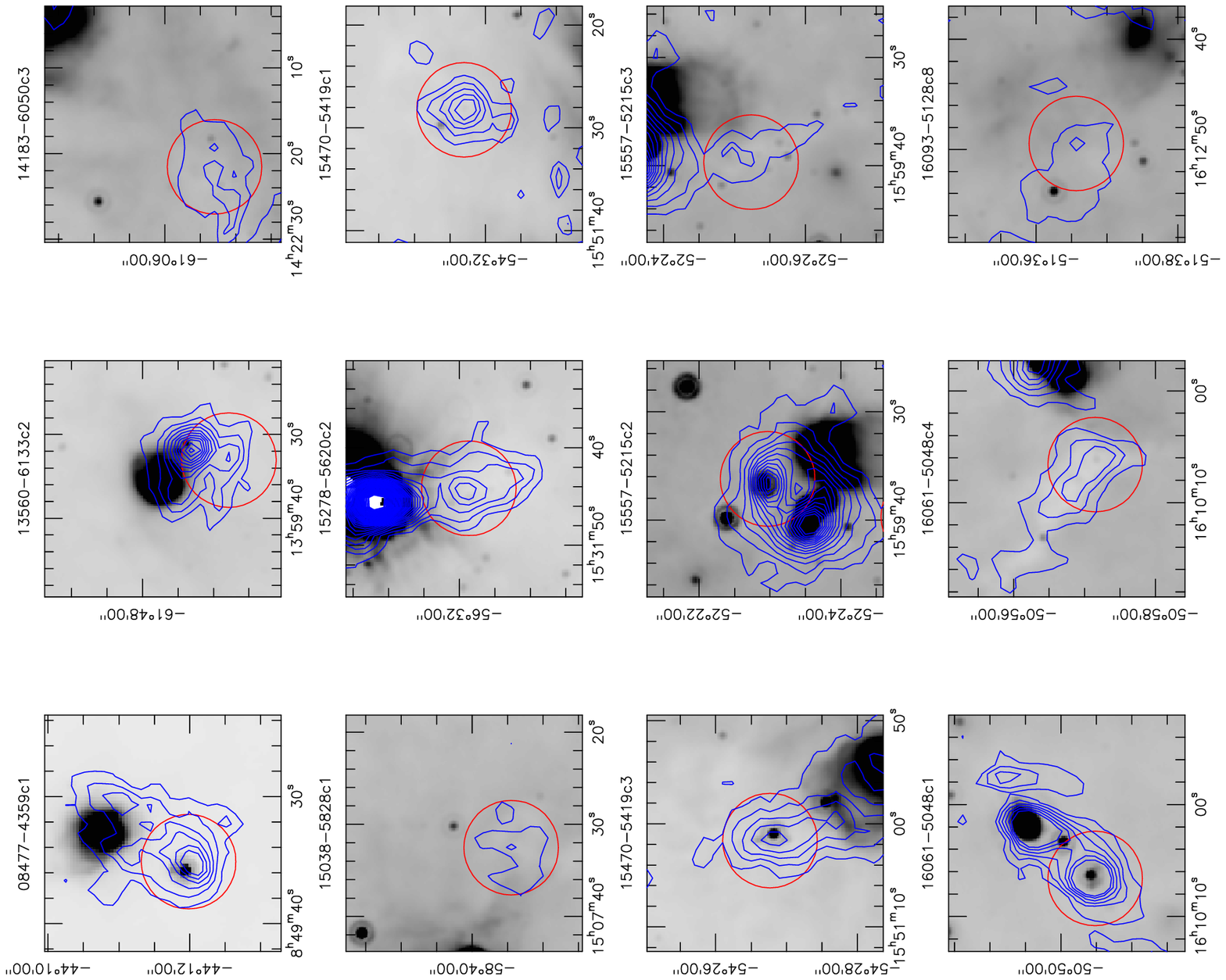}}
 \caption[]
 {\label{appA-fig1}{Superimposition of the Spitzer-MIPS 24 $\mu$m images
on the SIMBA 1.2~mm continuum maps (blue contours: first contour and
step correspond to the 3 $\sigma$ rms level)
towards the IRDCs studied in this work and observed
with Spitzer-MIPS (for clump 13039--6108c6, the Spitzer-MIPS image
is not available).
In each frame, the red circle indicates the targeted millimetre clump.
}}
 \end{center}
 \end{minipage}
\end{figure*}

\addtocounter{figure}{-1}
\begin{figure*}
\begin{minipage}{160mm}
 \begin{center}
 \resizebox{\hsize}{!}{\includegraphics[angle=-90]{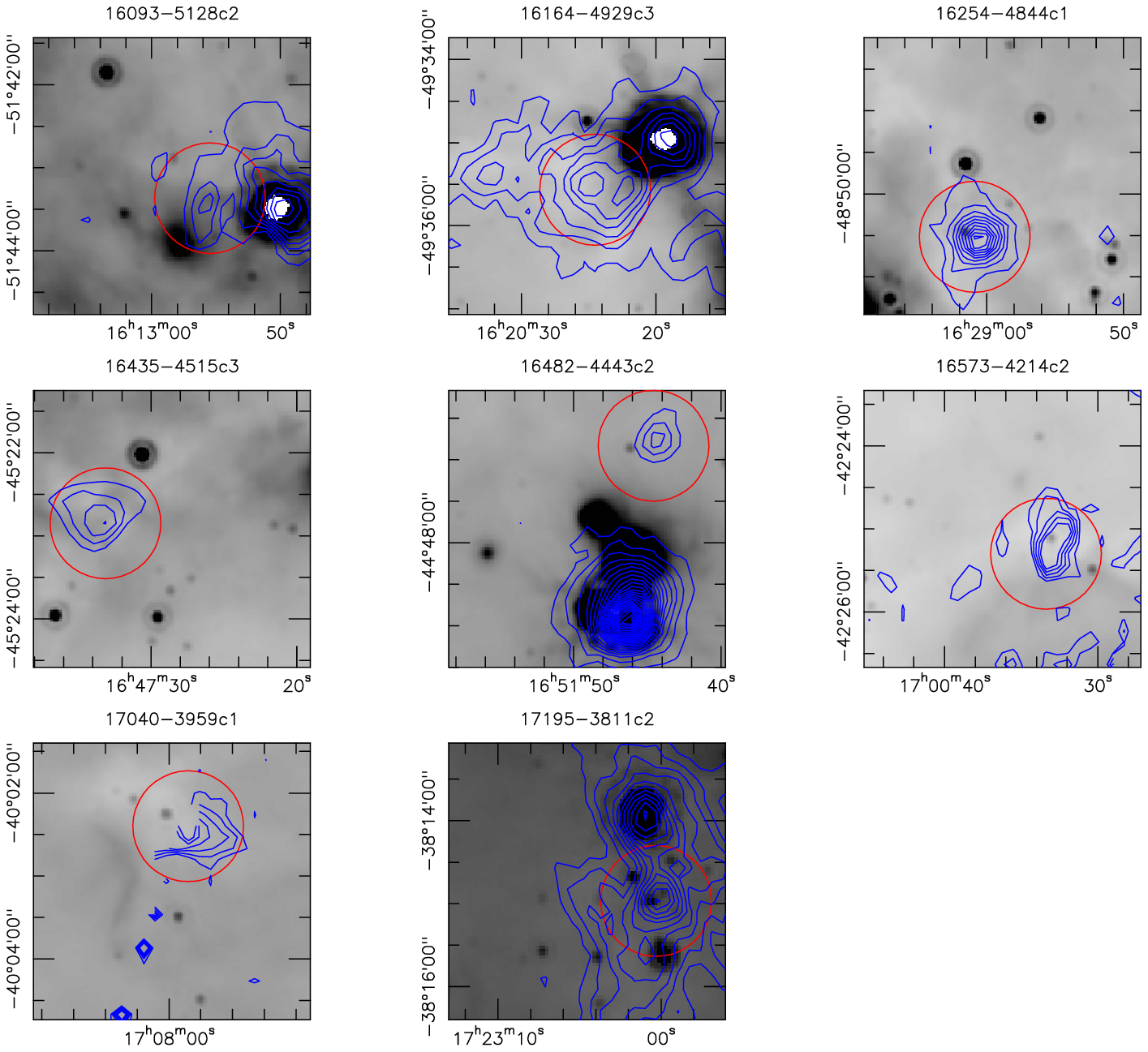}}
 \caption[]
 {\label{appA-fig2}{Continued.}}
 \end{center}
 \end{minipage}
\end{figure*} 

\section[]{Spectra}
\label{spectra}

\begin{figure*}
\begin{minipage}{160mm}
 \begin{center}
 \resizebox{\hsize}{!}{\includegraphics[angle=0]{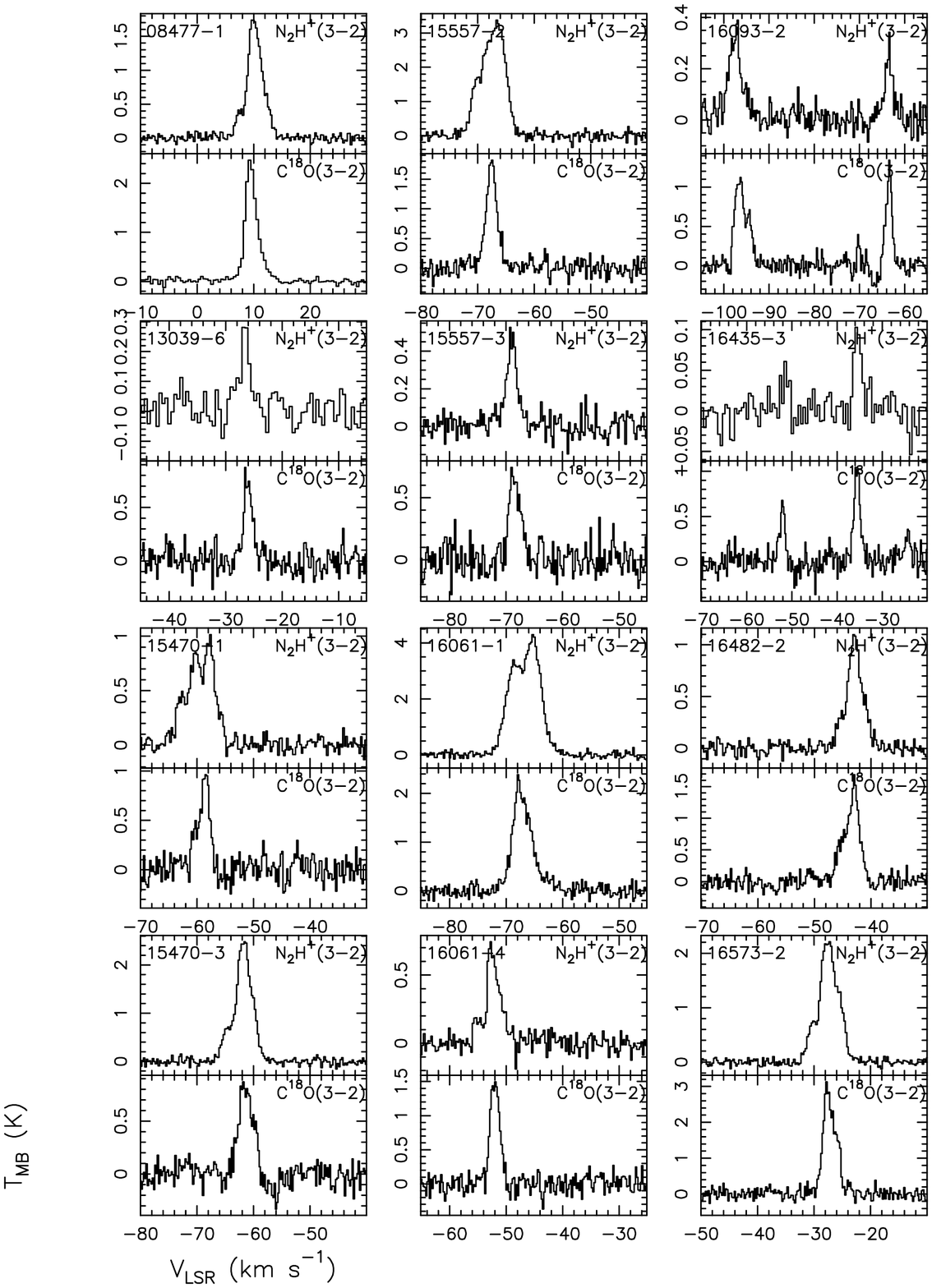}}
 \caption[]
 {\label{spectra_co_n2hp}{Spectra of the twelve sources observed and detected
 in both \H\ (3--2) and \CII\ (3--2). In each frame, we show a range of $\pm 20$ \kms\
around the systemic velocity adopted to centre the spectra (given in 
Table~\ref{sources}) and derived from CS observations.
For two objects, 16093--5128c2 and 16435--4515c3, the spectrum contains 
velocity components separated by several \kms , so that we show a broader velocity range. }}
 \end{center}
 \end{minipage}
\end{figure*}

\begin{figure*}
\begin{minipage}{160mm}
 \begin{center}
 \resizebox{\hsize}{!}{\includegraphics[angle=-90]{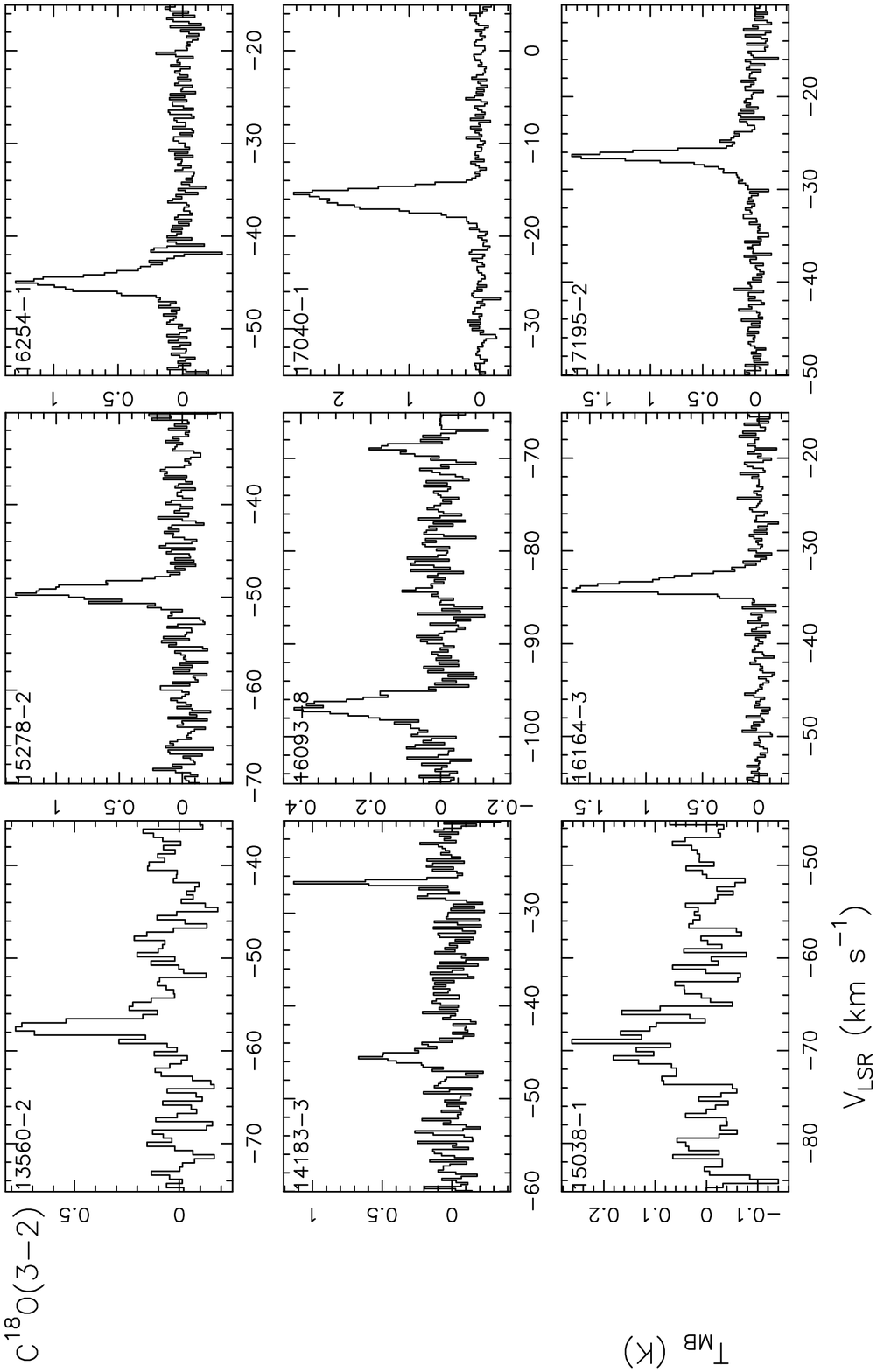}}
 \caption[]
 {\label{spectra_co_only}{Spectra of the nine sources observed and detected
 in \CII\ (3--2) only. As in Fig.~\ref{spectra_co_n2hp}, the velocity interval on the x-axis range from
 \Vlsr\ -- 20 \kms\ to \Vlsr\ + 20 \kms .}}
 \end{center}
 \end{minipage}
\end{figure*}

\bsp

\label{lastpage}

\end{document}